\documentclass[twocolumn,showpacs,amsmath,amssymb,superscriptaddress,floatfix]{revtex4-2}
\usepackage[breaklinks=true,colorlinks,citecolor=blue,linkcolor=blue,urlcolor=blue]{hyperref}
\usepackage[T1]{fontenc}
\usepackage{lineno}
\usepackage{textcomp}
\usepackage{amsmath}
\usepackage{graphicx}
\usepackage{multirow}
 \usepackage{makecell}
 \usepackage{booktabs}
\usepackage{derivative}
\usepackage{bbold}
\usepackage{lipsum}
\usepackage{array}
\usepackage{graphicx}
\usepackage{xcolor}
\usepackage{subfigure}
\usepackage{hyperref,hypcap}
\usepackage{braket}
\usepackage{amsmath}
\usepackage{commath}
\usepackage{mathtools}
\usepackage{chemformula}
\usepackage{natbib}
\usepackage{comment}
\usepackage{todonotes}
\newcolumntype{P}[1]{>{\centering\arraybackslash}p{#1}}
\usepackage{orcidlink}

\def\be{\begin{equation}}
\def\ee{\end{equation}}

\def\bea{\begin{eqnarray}}
\def\eea{\end{eqnarray}}

\makeatletter
\usepackage{lipsum}
\usepackage{epsfig}

\usepackage[colorlinks,linkcolor=blue]{hyperref}
\usepackage[nameinlink,capitalise]{cleveref}

\usepackage{subfigure}
\usepackage{color}
\usepackage{physics}

\definecolor{Red}{rgb}{1,0,0}
\definecolor{Blu}{rgb}{0,0,1}
\definecolor{Green}{rgb}{0,1,0}

\makeatother
\usepackage{amssymb}
\usepackage[english]{babel}
\begin{document}

\title{Quantum-Geometric Subgap Second-Harmonic Generation 
in $\mathcal{PT}$-Symmetric Antiferromagnetic Insulators}

\author{Shakeel Ahmad\orcidlink{0009-0000-4994-9717}}
\affiliation{Department of Physics, University of Alabama at Birmingham, Birmingham 35294, AL, USA}

\author{Fei Xue\,\orcidlink{0000-0002-1737-2332}} 
\affiliation{Department of Physics, University of Alabama at Birmingham, Birmingham 35294, AL, USA}

\begin{abstract}
Although quantum geometry plays an established role in
electronic transport and optical responses, the contribution 
of the time-reversal-odd Riemannian connection to fully 
subgap responses remains less explored. Here we identify intrinsic second-harmonic
generation in $\mathcal{PT}$-symmetric antiferromagnets,
predominantly governed by the real sector of the interband
Hermitian connection and arising entirely from virtual
interband transitions. In bilayer MnBi$_2$Te$_4$, the
response persists in trivial phases but is enhanced in
the layer-Chern phase and reverses sign with N\'eel order,
offering a probe of quantum geometry and phase-sensitive
readout of antiferromagnetic order without mobile carriers
or real excitations.
\end{abstract}
\date{\today}
\maketitle

\textit{Introduction}---
The geometry of Bloch wave functions plays a central role in electronic
transport and optical response. The Berry curvature governs the intrinsic
anomalous Hall effect and underlies Chern insulating phases~\cite{Niu2010,Nagaosa2010,Review_QAHE}. Beyond linear
response, nonlinear transport provides access to additional geometric quantities, including the Berry-curvature dipole, quantum metric, and higher-order Berry connections~\cite{Gao2014,Sodeman2015,Xu2018,Ma2019,Xiao2021,Yang2021,Das2023,Gao2023,Mehraeen2024}. Linear and nonlinear optical responses likewise encode interband quantum geometry through the quantum metric and multiband connections
~\cite{Nagaosa2016,Holder2020,Hsieh2021,Ahn2022,Refael2022,Song2023,
Fiete2024,Bansil2024,Wu2025,Moore2025,He2026,Claassen2026,Froese2026}. Together with
recent broader developments and reviews
~\cite{Jiang_2025,Yu2025,Xie2025,Queiroz2026,Review_QuantumGeometry}, these works establish
nonlinear response as a powerful probe of the geometric structure of Bloch states.

The geometric mechanisms underlying nonlinear response depend strongly on the excitation regime. Nonlinear Hall transport commonly involves Fermi-surface carriers, whereas shift and injection currents require real interband excitations~\cite{Sodeman2015,Sipe2000,
Souza2018,Ahn2022}. Recent studies of quantum-geometric second-harmonic generation (SHG) have also emphasized resonant optical transitions~\cite{Bhalla2022,Okyay2022,
He2026}. Although off-resonant SHG in insulators is
well established~\cite{Aversa1995,Hughes1996},
its geometric interpretation has attracted renewed
interest, including studies of quantum-metric-enhanced SHG in the transparent regime of a Haldane
model~\cite{Wang2025}.
In the fully subgap regime, $2\hbar\omega<E_{\rm g}$,
neither one- nor two-photon interband absorption is
possible, and the response arises entirely from
virtual interband processes. What geometric structures govern this intrinsic response in magnetic insulators, and how does magnetic symmetry constrain its low-frequency behavior?

Here we establish a quantum-geometric description
of fully subgap SHG in $\mathcal{PT}$-symmetric
antiferromagnets. We identify the real (Riemannian)
sector ~\cite{Ahn2022,Hetenyi2023} of the interband Hermitian connection as a
key contribution, together with genuine three-band
geometric terms. The Riemannian sector is odd under
time reversal, allowing a magnetism-sensitive nonlinear response even when $\mathcal{PT}$ symmetry forbids the intrinsic anomalous Hall effect. Using bilayer MnBi$_2$Te$_4$, an intrinsic antiferromagnetic topological insulator platform with layer-resolved topology~\cite{Otrokov2019,Gao2021}, in which strong nonlinear
optical responses have previously been
predicted~\cite{Fei2020}, we demonstrate that magnetic symmetry controls the leading low-frequency behavior of the in-plane SHG conductivity. The response begins at
quartic order in frequency for out-of-plane N\'eel order, whereas rotating the N\'eel vector permits a quadratic contribution. Remarkably, the subgap response is strongly enhanced in the layer-Chern phase relative to topologically trivial phases with the same direct band gap, highlighting
the role of Bloch-state geometry beyond gap-size effects. Furthermore, reversing the N\'eel order reverses the SHG amplitude, offering a phase-sensitive route to distinguishing opposite antiferromagnetic states in the fully insulating,
off-resonant regime, without requiring Fermi-surface carriers or real interband excitations. This provides a complementary
approach to earlier nonlinear transport methods for
N\'eel-vector detection ~\cite{Godinho2018,Xiao2021,Yang2021,
Gao2023,Wang2023_Experiment}.

\begin{figure}[htbp]
    \includegraphics[width=0.9\linewidth]{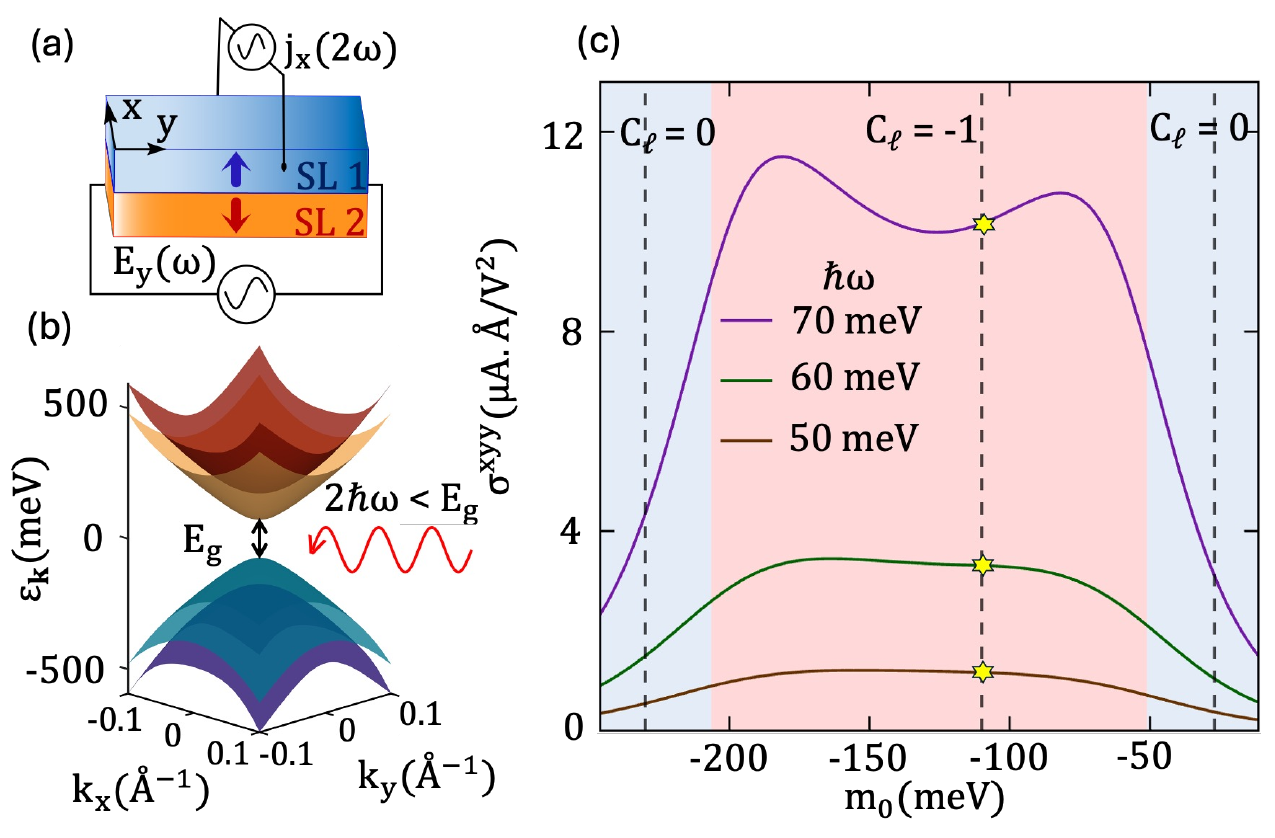}
    \caption{
    (a) Oppositely magnetized septuple layers (SLs) driven by
    $E_y(\omega)$, inducing a transverse current $j_x(2\omega)$.
    (b) Calculated band structure near $\Gamma$, showing the
    insulating gap $E_{\rm g}$. For $2\hbar\omega<E_{\rm g}$, SHG
    arises entirely from virtual interband transitions.
    (c) SHG conductivity $\sigma^{xyy}$ versus the band-inversion
    parameter $m_0$ at $\hbar\omega=50$, $60$, and $70~\mathrm{meV}$.
    Shading denotes the $C_\ell=0$ and $C_\ell=-1$ phases.
    Stars mark the reference layer-Chern state
    ($m_0^*=-110~\mathrm{meV}$); the outer dashed lines
    mark two trivial states with the same direct gap but
    substantially weaker SHG responses.
    }
    \label{fig:figure1}
\end{figure}

\textit{Intrinsic subgap second-harmonic conductivity}---
We specialize the length-gauge density-matrix formulation of
Ref.~\cite{Ahmad2026} to second-harmonic generation (SHG) in
a gapped insulator. With
$H(t)=H_0-qE^b(t)x^b$ and
$j^a(t)=q\,\mathrm{Tr}[v^a\rho(t)]$,
where $q=-e$ and
$v^a=(i/\hbar)[H_0,x^a]$,
we adopt the frequency-domain convention of
Ref.~\cite{Ahmad2026} for two equal incident
frequencies, $\omega_1=\omega_2=\omega$:
\begin{equation}
j^{a(2)}(2\omega)
=
\sigma^{abc}(2\omega;\omega,\omega)
E^b(\omega)E^c(\omega).
\label{eq:shg_definition}
\end{equation}
Here $E^b(\omega)$ is the positive-frequency
amplitude defined by
$E^b(t)=E^b(\omega)e^{-i\omega t}+\mathrm{c.c.}$.
The corresponding real time-dependent SHG current is
$j_{\rm SHG}^a(t)=2\operatorname{Re}\left[j^{a(2)}(2\omega)e^{-2i\omega t}\right].$
We consider the fully subgap regime
$2\hbar\omega<E_{\rm g}$, with the chemical
potential inside the insulating gap, where SHG
arises entirely from virtual interband transitions
[Figs.~\ref{fig:figure1}(a) and \ref{fig:figure1}(b)].

Expanding the density matrix to second order in the electric field and
combining its diagonal and off-diagonal sectors gives
\begin{multline}
\sigma^{abc}(2\omega;\omega,\omega)
=
\frac{q^3}{\hbar^2}
\sum_{m\neq n}
\Bigg\{
\frac{
f_{mn}\Delta^a_{nm}r^b_{nm}r^c_{mn}
}{
\omega_{mn}^2-\omega^2
}
\\
+
\bigg[
F_{mn}(\omega)r^{b;a}_{nm}r^c_{mn}
+
B_{mn}(2\omega)F_{mn}(\omega)
r^{a;b}_{nm}r^c_{mn}
\\
\qquad
+iB_{mn}(2\omega)r^a_{nm}
\big[r^b,F(\omega)r^c\big]'_{mn}
\\
\qquad
+
F_{mn}(\omega)
\big[\partial_bB_{mn}(2\omega)\big]
r^a_{nm}r^c_{mn}
+
(b\leftrightarrow c)
\bigg]
\Bigg\}.
\label{eq:intrinsic_shg_conductivity}
\end{multline}
Here
$\omega_{mn}\equiv(\varepsilon_m-\varepsilon_n)/\hbar$, $f_{mn}\equiv f_m-f_n$ where
$f_n=f(\varepsilon_n)$ is the Fermi-Dirac occupation, and
$\Delta^a_{nm}\equiv\partial_a\omega_{nm}
=v^a_{nn}-v^a_{mm}$, with $\partial_a=\partial/\partial k_a$.
For brevity, we write
$F_{mn}(\omega)=f_{mn}/(\omega_{mn}-\omega)$
and
$B_{mn}(2\omega)=2\omega/(\omega_{mn}-2\omega)$.
The interband Berry connection and its generalized derivative ($n\neq m$) are
\begin{equation}
    r^a_{nm}
    =
    i\langle u_n|\partial_a u_m\rangle; 
    r^{a;b}_{nm}
    =
    \partial_b r^a_{nm}
    -i(A^b_{nn}-A^b_{mm})r^a_{nm},
\end{equation}
where $A^b_{nn}=i\langle u_n|\partial_bu_n\rangle$.
The primed commutator contains genuine three-band
processes,
\begin{equation}
    \big[r^b,F(\omega)r^c\big]'_{mn}
    =
    \sum_{\ell\neq m,n}
    \left[
    r^b_{m\ell}F_{\ell n}(\omega)r^c_{\ell n}
    -
    F_{m\ell}(\omega)r^c_{m\ell}r^b_{\ell n}
    \right].
    \label{eq:three_band_commutator}
\end{equation}
Band summations include momentum integration, and
$(b\leftrightarrow c)$ applies only to the square brackets.
The first term in
Eq.~\eqref{eq:intrinsic_shg_conductivity}
comes from the diagonal density matrix, while the remaining
terms describe interband coherence.
A detailed derivation is given in the Supplemental Material.

In the static limit, the insulating occupations satisfy
$\partial_a f_n=0$, and
Eq.~\eqref{eq:intrinsic_shg_conductivity}
reduces to a Brillouin-zone total derivative,
\begin{equation}
    \sigma^{abc}(0;0,0)
    =
    \frac{q^3}{\hbar^2}
    \sum_{m\neq n}
    \partial_a
    \left[
        \frac{f_{mn}}{\omega_{mn}}
        r^b_{nm}r^c_{mn}
    \right]
    =0.
    \label{eq:static_zero}
\end{equation}
Thus, the intrinsic second order current vanishes in the static limit of a
gapped insulator. At finite frequency this cancellation is incomplete,
leaving a purely dynamical response generated by virtual interband
transitions.

The geometric content of Eq.~\eqref{eq:intrinsic_shg_conductivity}
is revealed by the transition Hermitian metric and
connection~\cite{Ahn2022},
\begin{equation}
    Q^{mn}_{ba}
    \equiv
    r^b_{nm}r^a_{mn},
    \qquad
    \mathcal{C}^{mn}_{bca}
    \equiv
    r^b_{nm}r^{a;c}_{mn}.
    \label{eq:metric_connection}
\end{equation}
The real part of $Q^{mn}_{ba}$ defines the transition quantum metric,
while $\mathcal{C}^{mn}_{bca}$ describes the associated Hermitian
connection. In a multiband system, the connection can possess nonzero torsion,
\begin{equation}
    \mathcal{T}^{mn}_{bca}
    \equiv
    \mathcal{C}^{mn}_{bca}
    -
    \mathcal{C}^{mn}_{bac}
    =
    i r^b_{nm}
    \sum_{\ell\neq m,n}
    \left(
        r^c_{m\ell}r^a_{\ell n}
        -
        r^a_{m\ell}r^c_{\ell n}
    \right).
    \label{eq:transition_torsion}
\end{equation}
Equation~\eqref{eq:intrinsic_shg_conductivity} can therefore
be organized in terms of interband geometry: products of
interband Berry connections define $Q^{mn}$, products
involving generalized derivatives define $\mathcal C^{mn}$,
and additional contributions arise from frequency-weighted
three-band virtual processes. For spinful
$\mathcal{PT}$-symmetric systems, the band indices are
understood to label distinct energy subspaces containing
degenerate states. The inter-subspace Berry connections
are matrices, their generalized derivatives take the
non-Abelian covariant form, and closed products are
traced over the corresponding subspaces.

To expose this structure more directly, we expand the subgap response in
the low-frequency limit $\omega\ll|\omega_{mn}|$. The static term
vanishes according to Eq.~\eqref{eq:static_zero}. In the clean subgap
regime, the energy denominators are real, and $\mathcal{PT}$ relates
interband geometric contributions to their complex conjugates at the
same momentum. The imaginary (symplectic) contributions therefore
cancel, leaving a real conductivity. Together with the reality
condition
$\sigma^{abc}(-2\omega;-\omega,-\omega)
=[\sigma^{abc}(2\omega;\omega,\omega)]^*$,
this eliminates all odd powers of frequency. The surviving even-order
coefficients probe the real (Riemannian) sector of the Hermitian
connection, together with symmetry-allowed multiband contributions.
The subgap conductivity therefore takes the form
\begin{equation}
\sigma^{abc}(2\omega;\omega,\omega)
=
\omega^2\mathcal G^{abc}_2
+\omega^4\mathcal G^{abc}_4
+\omega^6\mathcal G^{abc}_6+\cdots .
\label{eq:even_frequency_expansion}
\end{equation}
In the two-band limit, where the explicit multiband torsion is absent,
the coefficients can be organized entirely in terms of
energy-normalized Hermitian connections. At quadratic order,
\begin{equation}
\mathcal G^{abc}_{2,\mathcal{C}}
=
\frac{10q^3}{\hbar^2}
\sum_{m\neq n}
\frac{f_{mn}}{\omega_{mn}^3}
\left[
\mathrm{Re}\,\mathcal C^{mn}_{(abc)}
-\mathrm{Re}\,\mathcal C^{mn}_{abc}
\right],
\label{eq:low_frequency_geometry}
\end{equation}
where $\mathcal C^{mn}_{(abc)}$ denotes full symmetrization over the
Cartesian indices. The full multiband coefficients also contain explicit three-band terms, while the quartic two-band contribution carries stronger inverse-energy weighting,
$f_{mn}/\omega_{mn}^5$.
Detailed expressions through $O(\omega^4)$ are given
in the Supplemental Material.

The geometric sector selected here differs from that appearing in
conventional linearly polarized shift current
~\cite{Ahn2022,Sipe2000,Souza2018}. The Hermitian connection
contains real (Riemannian) and imaginary (symplectic) components.
The linearly polarized shift current is a resonant response that
probes the imaginary sector of $\mathcal C^{mn}$, which is even
under time reversal. By contrast, the even-frequency subgap
response of a $\mathcal{PT}$-symmetric antiferromagnet arises
from virtual interband transitions and probes the real,
Riemannian sector. This distinction follows from
\begin{equation}
    \mathcal{C}_{abc}(\mathbf{k})
    \xrightarrow{\mathcal T}
    -\mathcal{C}_{abc}^{*}(-\mathbf{k}),
    \qquad
    \mathcal{C}_{abc}(\mathbf{k})
    \xrightarrow{\mathcal{PT}}
    \mathcal{C}_{abc}^{*}(\mathbf{k}).
\end{equation}
Thus, $\mathrm{Re}\,\mathcal C_{abc}$ is odd under
$\mathcal T$ but allowed by $\mathcal{PT}$, whereas
$\mathrm{Im}\,\mathcal C_{abc}$ has the opposite symmetry
character. In a noncentrosymmetric magnet without
$\mathcal{PT}$ symmetry, both geometric sectors can in
principle contribute to SHG.
The Riemannian subgap response considered here therefore
provides a symmetry-selective probe of quantum geometry
associated with broken time-reversal symmetry while
remaining compatible with $\mathcal{PT}$.

\textit{Topological $\mathcal{PT}$ antiferromagnet as an example}---
We apply our formalism to bilayer \ch{MnBi2Te4}, a collinear
antiferromagnet in which $\mathcal P$ and $\mathcal T$ are
individually broken while their product $\mathcal{PT}$ is
preserved. Its band-inverted electronic structure can support
nontrivial layer-Chern topology, providing a platform to
investigate how topology influences the intrinsic subgap SHG
response. Near the $\Gamma$ point, its low-energy electronic
structure is described by the eight-band $k\cdot p$
Hamiltonian~\cite{Xu2019,Cheng2024, Han2025}
\begin{equation}
\begin{split}
H_0&(\mathbf{k})={}
\mu_0\otimes[h_o(\mathbf{k})+h_w(\mathbf{k})]
+\mu_y\otimes A_1\sigma_z\otimes\tau_x\\
&-\mu_x\otimes
[D_1\sigma_0\otimes\tau_0+B_1\sigma_0\otimes\tau_z]
+\mu_z\otimes h_m ,
\end{split}
\label{eq:MBT}
\end{equation}
where
\begin{equation}
\begin{aligned}
h_o(\mathbf{k})
&=
d_0(\mathbf{k})\sigma_0\otimes\tau_z
+A_2(k_x\sigma_x+k_y\sigma_y)\otimes\tau_x,\\
h_m
&=
\lambda_{\rm exc}\sigma_z\otimes(m_+\tau_0+m_-\tau_z),\\
h_w
&=
-i\lambda_w(k_+^3-k_-^3)\sigma_z\otimes\tau_x .
\end{aligned}
\label{eq:MBT_blocks}
\end{equation}
Here $k_\pm=k_x\pm ik_y$, and $\mu_i$, $\tau_i$, and $\sigma_i$ act
in layer, orbital, and spin space, respectively. The staggered exchange
term $\mu_z\otimes h_m$ breaks $\mathcal P$ and $\mathcal T$
separately while preserving $\mathcal{PT}$, whereas $h_w$ reduces the
emergent continuous in-plane rotational symmetry to the crystalline
$C_{3z}$ symmetry. The band-inversion parameter $m_0$, entering
$d_0(\mathbf{k})=m_0+2B_1+B_2(k_x^2+k_y^2)$,
tunes the system between layer-Chern and topologically
trivial phases. 
The numerical values of the model parameters used in our calculations
are provided in the Supplemental Material.

For out-of-plane N\'eel order, the bilayer preserves
$C_{3z}$ and $\mathcal M_x\mathcal T$, in addition to
$\mathcal{PT}$. The combined $\mathcal{PT}$ and
$\mathcal M_x\mathcal T$ symmetries imply a unitary
twofold rotation $C_{2x}$, which requires
\begin{equation}
\sigma^{yyy}=\sigma^{yxx}
=\sigma^{xyx}=\sigma^{xxy}=0.
\label{eq:MBT_MxT}
\end{equation}
The $C_{3z}$ symmetry further relates the surviving
components according to~\cite{Mei2024}
\begin{equation}
\sigma^{xyy}
=
\sigma^{yxy}
=
\sigma^{yyx}
=
-\sigma^{xxx}.
\label{eq:MBT_tensor_symmetry}
\end{equation}
Thus, the in-plane SHG conductivity has a single independent
component, which we take to be $\sigma^{xyy}$, describing
the transverse current $j_x(2\omega)$ induced by
$E_y(\omega)$ [Fig.~\ref{fig:figure1}(a)]. 

We now examine how this response evolves with band inversion
and layer-Chern topology. Figure~\ref{fig:figure1}(b) 
shows the calculated band structure near $\Gamma$ for the 
reference ground-state parameters, while Fig.~\ref{fig:figure1}(c) 
shows the subgap SHG conductivity $\sigma^{xyy}$ versus the
band-inversion parameter $m_0$ at several photon energies.
The three dashed lines identify the reference layer-Chern
state and two neighboring topologically trivial states
with the same direct band gap, enabling a direct comparison
of SHG responses across distinct topological phases at
fixed gap. The response increases in magnitude with photon
energy, reflecting its dynamical origin. Although finite
in both topological and trivial insulating phases,
the conductivity is substantially enhanced over much
of the layer-Chern phase ($C_\ell=-1$) compared with the
neighboring trivial phases ($C_\ell=0$).

\begin{figure}[htbp]
    \centering
    \includegraphics[width=1\linewidth]{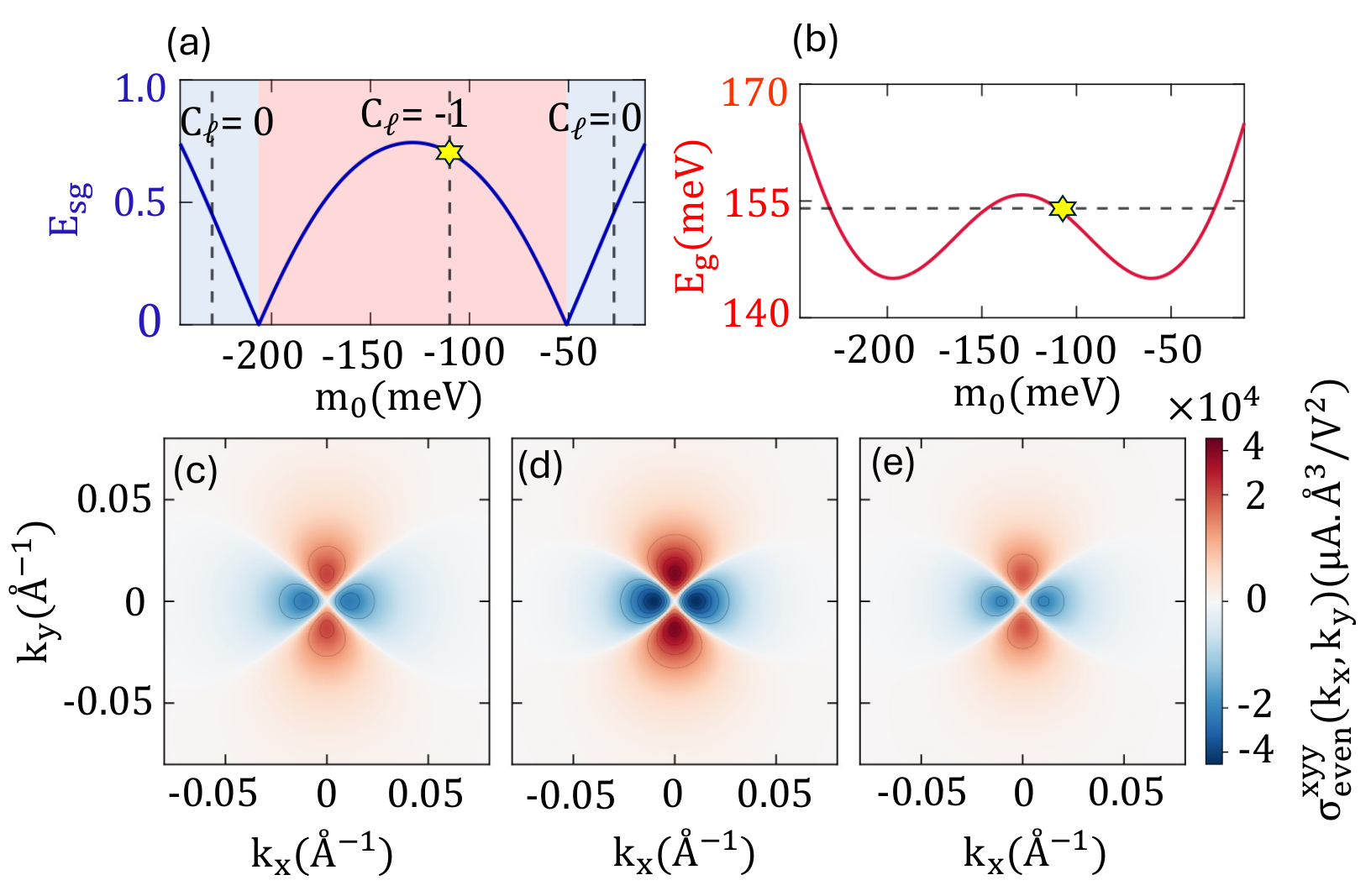}
    \caption{
    (a) Spectral gap $E_{\rm sg}$ of the projected layer-pseudospin
    operator $P(\mathbf{k})\mu_zP(\mathbf{k})$ versus the
    band-inversion parameter $m_0$. Spectral gap closings separate the
    $C_\ell=0$ and $C_\ell=-1$ phases.
    (b) Electronic band gap $E_{\rm g}$ versus $m_0$.
    Stars in (a) and (b) mark the reference layer-Chern state
    ($m_0=-110~\mathrm{meV}$); the horizontal dashed line
    indicates its gap ($E_{\rm g}\simeq154~\mathrm{meV}$).
    (c)--(e) Momentum-resolved $k_x$-even SHG contributions
    $\sigma^{xyy}_{\rm even}(\mathbf{k})$ at
    $m_0=-230$, $-110$, and $-27~\mathrm{meV}$,
    respectively (vertical dashed lines in (a)).
    Despite the same gap, the layer-Chern state (d)
    exhibits substantially stronger contributions
    than the trivial states (c) and (e).
    }
    \label{fig:figure2}
\end{figure}

To understand the enhanced SHG response in the layer-Chern
phase, we first characterize its layer-resolved topology.
Although $\mathcal{PT}$ symmetry forces the total Chern
number to vanish, the occupied manifold can retain
nontrivial layer-resolved topology. Following the
projected-pseudospin construction of
Prodan~\cite{Prodan2009,Prodan2011}, we diagonalize
the projected layer operator
$P(\mathbf{k})\mu_zP(\mathbf{k})$ within the occupied
subspace, where $P(\mathbf{k})$ is the occupied-band
projector. When a finite spectral gap separates the
positive and negative pseudospin sectors, each can be
assigned a Chern number $C_\pm$, defining the layer
Chern number $C_\ell=(C_+-C_-)/2$.
As shown in Fig.~\ref{fig:figure2}(a), the projected-pseudospin
gap closes at the boundaries of the $C_\ell=-1$ phase,
while the electronic gap remains finite
[Fig.~\ref{fig:figure2}(b)]. Thus, the layer-Chern
topological transition occurs through a closing of the
projected-pseudospin gap without closing the electronic gap.

Having established the topological distinction, we next
examine where the SHG enhancement originates in momentum
space. The momentum-resolved conductivity
$\sigma^{xyy}(\mathbf{k})$ contains a large contribution
that is odd under $k_x\rightarrow-k_x$ and cancels upon
momentum integration. To isolate the part contributing
to the net conductivity, we consider its $k_x$-even part,
\begin{equation}
\sigma^{xyy}_{\rm even}(k_x,k_y)
=
\frac{1}{2}
\left[
\sigma^{xyy}(k_x,k_y)
+
\sigma^{xyy}(-k_x,k_y)
\right],
\end{equation}
whose momentum integral equals that of the full
$\sigma^{xyy}(\mathbf{k})$.
Figures~\ref{fig:figure2}(c)-(e) compare three
representative states with the same direct electronic gap.
In the layer-Chern phase, the contributing weight is
strongly enhanced and concentrated near $\Gamma$,
where the low-energy bands are inverted, whereas it
is substantially weaker in the neighboring trivial
phases. Because the three states have the same direct
gap, this enhancement suggests a connection to the
interband geometry of the inverted bands.

To examine this connection more directly, we turn to
the low-frequency response, where the interplay between
quantum geometry and crystalline symmetry imposes an
additional selection rule. As shown in the Supplemental
Material, the complete quadratic coefficient, including
genuine three-band processes, has a vanishing fully
symmetric Cartesian part, $\mathcal G_2^{(abc)}=0$.
In particular, $\mathcal G_2^{xxx}=0$, and the
$C_{3z}$-constrained tensor relations in
Eq.~\eqref{eq:MBT_tensor_symmetry} then require all
in-plane components to vanish at quadratic order.
Since the static response vanishes and $\mathcal{PT}$
symmetry eliminates odd powers of frequency, the
leading in-plane subgap SHG response is quartic,
\begin{equation}
\sigma^{xyy}(2\omega;\omega,\omega)
=
\omega^4\mathcal G^{xyy}_4
+O(\omega^6).
\label{eq:MBT_omega4}
\end{equation}

Because the surviving in-plane SHG tensor is fully symmetric
under Cartesian-index permutations, the Hermitian-connection
contribution to the leading quartic coefficient simplifies to
\begin{equation}
\mathcal G^{xyy}_{4,\mathcal C}
=
-\frac{84q^3}{5\hbar^2}
\sum_{m\neq n}
\frac{f_{mn}}{\omega_{mn}^{5}}
\,\mathrm{Re}\,\mathcal C^{mn}_{(xyy)} ,
\label{eq:MBT_G4_connection}
\end{equation}
while the full coefficient also contains genuine three-band
contributions,
$\mathcal G^{xyy}_4=
\mathcal G^{xyy}_{4,\mathcal C}+
\mathcal G^{xyy}_{4,3{\rm b}}$.

\begin{figure}[htbp]
    \includegraphics[width=0.9\linewidth]{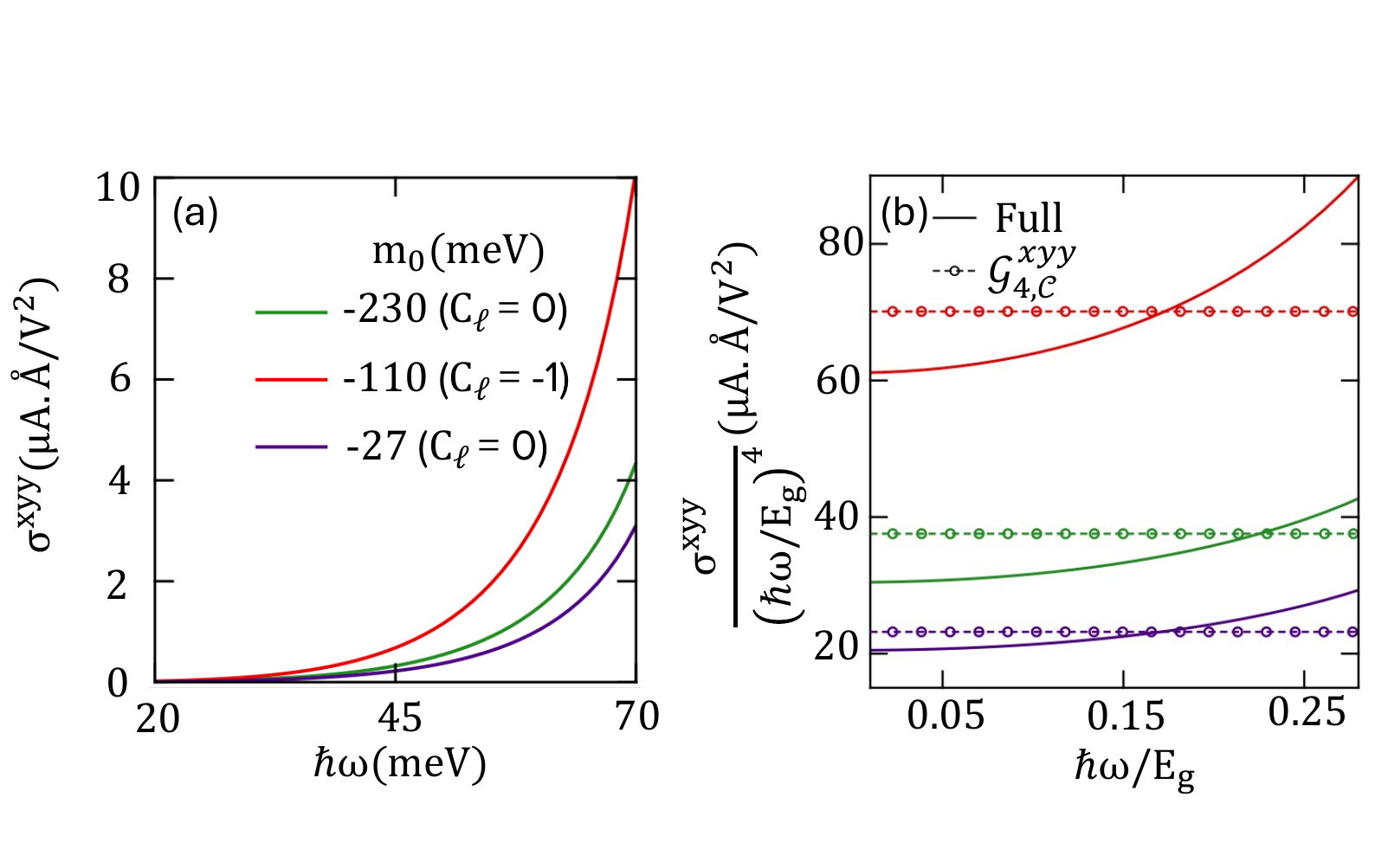}
    \caption{
    (a) Frequency dependence of $\sigma^{xyy}$ for three
    equal-gap states at $m_0=-230$, $-110$, and
    $-27~\mathrm{meV}$, corresponding to
    $C_\ell=0$, $-1$, and $0$, respectively.
    (b) Conductivity normalized by $(\hbar\omega/E_{\rm g})^4$,
    revealing the low-frequency plateaus characteristic
    of quartic scaling. Solid curves show the full response;
    horizontal dashed lines show the Hermitian-connection
    contribution $(E_{\rm g}^4/\hbar^4)\mathcal G^{xyy}_{4,\mathcal C}$
    from Eq.~\eqref{eq:MBT_G4_connection}, excluding explicit
    three-band terms. The geometric contribution closely
    approximates the full low-frequency response and is
    substantially enhanced in the layer-Chern phase.
    }
    \label{fig:figure3}
\end{figure}
Figure~\ref{fig:figure3}(a) shows the frequency dependence
of $\sigma^{xyy}$ for the three equal-gap states, with a
substantially stronger response in the layer-Chern phase.
To examine the predicted quartic scaling at low frequency,
Fig.~\ref{fig:figure3}(b) shows the conductivity normalized
by $(\hbar\omega/E_{\rm g})^4$, which approaches
$(E_{\rm g}^4/\hbar^4)\mathcal G^{xyy}_4$ as $\omega\to0$.
All three responses exhibit low-frequency plateaus,
confirming the quartic onset predicted by
Eq.~\eqref{eq:MBT_omega4}.
Because the three states have the same $E_{\rm g}$, these
plateaus directly compare their quartic coefficients.
The horizontal dashed lines in Fig.~\ref{fig:figure3}(b) 
represent the corresponding Hermitian-connection contributions
$(E_{\rm g}^4/\hbar^4)\mathcal G^{xyy}_{4,\mathcal C}$,
which provide a good approximation to the full
low-frequency coefficients, capturing both their
relative magnitudes and the pronounced enhancement
in the layer-Chern phase. The remaining differences
arise from genuine three-band processes, while deviations
at higher frequencies also reflect higher-order terms.
These results identify the inverse-energy-weighted
Riemannian connection as a major contribution to the
enhanced subgap SHG response.

\begin{figure}[b]
    \centering
    \includegraphics[width=0.85\linewidth]{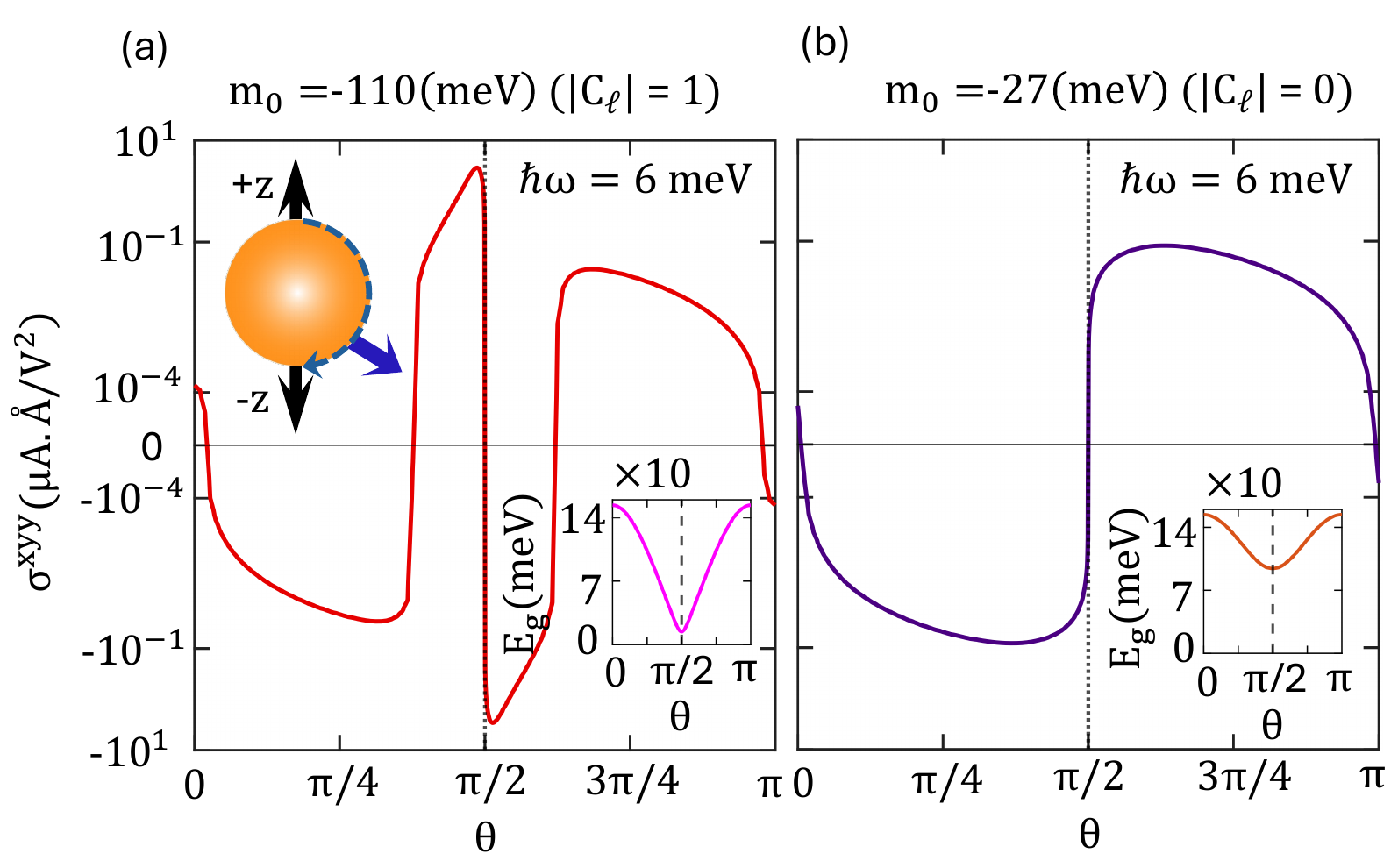}
    \caption{
    Second-harmonic conductivity $\sigma^{xyy}$ versus the
    N\'eel-vector angle $\theta$ for
    (a) $m_0=-110~\mathrm{meV}$ ($C_\ell=-1$) and
    (b) $m_0=-27~\mathrm{meV}$ ($C_\ell=0$) with $\hbar\omega=6$~meV.
    The N\'eel vector rotates in the $xz$ plane, with
    $\theta=0$ and $\pi$ corresponding to opposite
    out-of-plane orientations.
    The response reverses sign between these states and
    vanishes at $\theta=\pi/2$ by mirror symmetry
    (vertical dotted lines).
    Additional sign reversals near the out-of-plane
    configurations arise from competing quadratic and
    quartic contributions.
    Insets show the corresponding electronic gap
    $E_{\rm g}(\theta)$. 
    }
    \label{fig:figure4}
\end{figure}

Beyond its quantum-geometric origin, fully subgap SHG
offers a promising route to probing N\'eel order in
compensated antiferromagnetic insulators, where
$\mathcal{PT}$ symmetry forbids conventional anomalous
Hall readout. To examine its sensitivity to magnetic
orientation, we rotate the N\'eel vector in the $xz$
plane according to
$\mathbf n(\theta)=(\sin\theta,0,\cos\theta)$.
Tilting the N\'eel vector breaks $C_{3z}$ symmetry,
allowing a quadratic $\omega^2$ contribution that is
forbidden for out-of-plane order, while $\mathcal{PT}$
symmetry continues to eliminate odd powers of frequency.

The angular dependence is further constrained by mirror
symmetry. Under $\mathcal M_x$, an axial vector 
transforms as $(n_x,n_y,n_z)\rightarrow(n_x,-n_y,-n_z)$,
mapping $\mathbf n(\theta)$ to
$\mathbf n(\pi-\theta)$.
Since $j_x$ is odd while $E_y^2$ is even under
$\mathcal M_x$, the conductivity satisfies
$\sigma^{xyy}(\theta)=-\sigma^{xyy}(\pi-\theta)$.
This relation enforces a vanishing response at
$\theta=\pi/2$ and opposite SHG signs for the two
out-of-plane N\'eel states, enabling their discrimination
through phase-sensitive detection.
Figure~\ref{fig:figure4} demonstrates a pronounced
sensitivity to N\'eel-vector rotation, including sign
reversals near the out-of-plane configurations.
For the parameters considered, these additional zeros
arise from competition between the rotation-induced
quadratic term and the quartic term, which have opposite
signs. Unlike the symmetry-enforced zero at $\theta=\pi/2$,
these additional sign reversals depend on the electronic
structure and driving frequency.
Together, the symmetry-enforced N\'eel-state contrast
and the strong sensitivity to small rotations highlight
fully subgap SHG as a promising probe of antiferromagnetic
order without requiring mobile carriers or real
interband excitations.

\textit{Conclusion}---
In summary, we establish an intrinsic, fully subgap
second-harmonic response in $\mathcal{PT}$-symmetric
antiferromagnetic insulators, arising entirely from
virtual interband transitions. The response probes the
time-reversal-odd Riemannian sector of the interband
Hermitian connection, with additional genuine multiband
contributions. In bilayer \ch{MnBi2Te4}, a geometric
selection rule combined with $C_{3z}$ symmetry produces
a quartic low-frequency response, whose magnitude is
strongly enhanced in the layer-Chern phase even at fixed
direct gap. Rotating the N\'eel vector permits a quadratic
contribution, while opposite N\'eel states exhibit a
symmetry-enforced sign contrast. These findings highlight
fully subgap SHG as a promising optical probe of both
the quantum-geometric Hermitian connection and N\'eel
order, without requiring mobile carriers or real
interband excitations.

{\it Acknowledgements}--- The work is supported by the National Science Foundation under Grant No. OIA-2229498 and UAB startup fund. 
We thank Farzad Mahfouzi, Paul Haney, and Mark Stiles for fruitful discussions.

\bibliography{references}{}

@PREAMBLE{
 "\providecommand{\noopsort}[1]{}"
 # "\providecommand{\singleletter}[1]{#1}%"
}

@article{Das2023,
  title = {Intrinsic nonlinear conductivities induced by the quantum metric},
  author = {Das, Kamal and Lahiri, Shibalik and Atencia, Rhonald Burgos and Culcer, Dimitrie and Agarwal, Amit},
  journal = {Phys. Rev. B},
  volume = {108},
  issue = {20},
  pages = {L201405},
  numpages = {6},
  year = {2023},
  month = {Nov},
  publisher = {American Physical Society},
  doi = {10.1103/PhysRevB.108.L201405},
  url = {https://link.aps.org/doi/10.1103/PhysRevB.108.L201405}
}

@article{Sipe2000,
  title = {Second-order optical response in semiconductors},
  author = {Sipe, J. E. and Shkrebtii, A. I.},
  journal = {Phys. Rev. B},
  volume = {61},
  issue = {8},
  pages = {5337--5352},
  numpages = {0},
  year = {2000},
  month = {Feb},
  publisher = {American Physical Society},
  doi = {10.1103/PhysRevB.61.5337},
  url = {https://link.aps.org/doi/10.1103/PhysRevB.61.5337}
}

@article{Aversa1995,
  title = {Nonlinear optical susceptibilities of semiconductors: Results with a length-gauge analysis},
  author = {Aversa, Claudio and Sipe, J. E.},
  journal = {Phys. Rev. B},
  volume = {52},
  issue = {20},
  pages = {14636--14645},
  numpages = {0},
  year = {1995},
  month = {Nov},
  publisher = {American Physical Society},
  doi = {10.1103/PhysRevB.52.14636},
  url = {https://link.aps.org/doi/10.1103/PhysRevB.52.14636}
}

@article{Nagaosa2010,
  title = {Anomalous Hall effect},
  author = {Nagaosa, Naoto and Sinova, Jairo and Onoda, Shigeki and MacDonald, A. H. and Ong, N. P.},
  journal = {Rev. Mod. Phys.},
  volume = {82},
  issue = {2},
  pages = {1539--1592},
  numpages = {0},
  year = {2010},
  month = {May},
  publisher = {American Physical Society},
  doi = {10.1103/RevModPhys.82.1539},
  url = {https://link.aps.org/doi/10.1103/RevModPhys.82.1539}
}

@article{Niu2010,
  title = {Berry phase effects on electronic properties},
  author = {Xiao, Di and Chang, Ming-Che and Niu, Qian},
  journal = {Rev. Mod. Phys.},
  volume = {82},
  issue = {3},
  pages = {1959--2007},
  numpages = {0},
  year = {2010},
  month = {Jul},
  publisher = {American Physical Society},
  doi = {10.1103/RevModPhys.82.1959},
  url = {https://link.aps.org/doi/10.1103/RevModPhys.82.1959}
}

@article{Yang2021,
  title = {Intrinsic Second-Order Anomalous Hall Effect and Its Application in Compensated Antiferromagnets},
  author = {Liu, Huiying and Zhao, Jianzhou and Huang, Yue-Xin and Wu, Weikang and Sheng, Xian-Lei and Xiao, Cong and Yang, Shengyuan A.},
  journal = {Phys. Rev. Lett.},
  volume = {127},
  issue = {27},
  pages = {277202},
  numpages = {6},
  year = {2021},
  month = {Dec},
  publisher = {American Physical Society},
  doi = {10.1103/PhysRevLett.127.277202},
  url = {https://link.aps.org/doi/10.1103/PhysRevLett.127.277202}
}

@article{Xiao2021,
  title = {Intrinsic Nonlinear Hall Effect in Antiferromagnetic Tetragonal CuMnAs},
  author = {Wang, Chong and Gao, Yang and Xiao, Di},
  journal = {Phys. Rev. Lett.},
  volume = {127},
  issue = {27},
  pages = {277201},
  numpages = {6},
  year = {2021},
  month = {Dec},
  publisher = {American Physical Society},
  doi = {10.1103/PhysRevLett.127.277201},
  url = {https://link.aps.org/doi/10.1103/PhysRevLett.127.277201}
}

@article{Gao2014,
  title = {Field Induced Positional Shift of Bloch Electrons and Its Dynamical Implications},
  author = {Gao, Yang and Yang, Shengyuan A. and Niu, Qian},
  journal = {Phys. Rev. Lett.},
  volume = {112},
  issue = {16},
  pages = {166601},
  numpages = {5},
  year = {2014},
  month = {Apr},
  publisher = {American Physical Society},
  doi = {10.1103/PhysRevLett.112.166601},
  url = {https://link.aps.org/doi/10.1103/PhysRevLett.112.166601}
}

@article{Wang2023_Experiment,
   author = {Wang, Naizhou and Kaplan, Daniel and Zhang, Zhaowei and Holder, Tobias and Cao, Ning and Wang, Aifeng and Zhou, Xiaoyuan and Zhou, Feifei and Jiang, Zhengzhi and Zhang, Chusheng and Ru, Shihao and Cai, Hongbing and Watanabe, Kenji and Taniguchi, Takashi and Yan, Binghai and Gao, Weibo},
   title = {Quantum-metric-induced nonlinear transport in a topological antiferromagnet},
   journal = {Nature},
   volume = {621},
   number = {7979},
   pages = {487-492},
   ISSN = {1476-4687},
   DOI = {10.1038/s41586-023-06363-3},
   url = {https://doi.org/10.1038/s41586-023-06363-3},
   year = {2023},
   type = {Journal Article}
}

@article{Sodeman2015,
  title = {Quantum Nonlinear Hall Effect Induced by Berry Curvature Dipole in Time-Reversal Invariant Materials},
  author = {Sodemann, Inti and Fu, Liang},
  journal = {Phys. Rev. Lett.},
  volume = {115},
  issue = {21},
  pages = {216806},
  numpages = {5},
  year = {2015},
  month = {Nov},
  publisher = {American Physical Society},
  doi = {10.1103/PhysRevLett.115.216806},
  url = {https://link.aps.org/doi/10.1103/PhysRevLett.115.216806}
}

@article{Review_QAHE,
  title = {Colloquium: Quantum anomalous Hall effect},
  author = {Chang, Cui-Zu and Liu, Chao-Xing and MacDonald, Allan H.},
  journal = {Rev. Mod. Phys.},
  volume = {95},
  issue = {1},
  pages = {011002},
  numpages = {33},
  year = {2023},
  month = {Jan},
  publisher = {American Physical Society},
  doi = {10.1103/RevModPhys.95.011002},
  url = {https://link.aps.org/doi/10.1103/RevModPhys.95.011002}
}

@article{Ma2019,
   author = {Ma, Qiong and Xu, Su-Yang and Shen, Huitao and MacNeill, David and Fatemi, Valla and Chang, Tay-Rong and Mier Valdivia, Andrés M. and Wu, Sanfeng and Du, Zongzheng and Hsu, Chuang-Han and Fang, Shiang and Gibson, Quinn D. and Watanabe, Kenji and Taniguchi, Takashi and Cava, Robert J. and Kaxiras, Efthimios and Lu, Hai-Zhou and Lin, Hsin and Fu, Liang and Gedik, Nuh and Jarillo-Herrero, Pablo},
   title = {Observation of the nonlinear Hall effect under time-reversal-symmetric conditions},
   journal = {Nature},
   volume = {565},
   number = {7739},
   pages = {337-342},
   ISSN = {1476-4687},
   DOI = {10.1038/s41586-018-0807-6},
   url = {https://doi.org/10.1038/s41586-018-0807-6},
   year = {2019},
   type = {Journal Article}
}

@article{Xu2018,
   author = {Xu, Su-Yang and Ma, Qiong and Shen, Huitao and Fatemi, Valla and Wu, Sanfeng and Chang, Tay-Rong and Chang, Guoqing and Valdivia, Andrés M. Mier and Chan, Ching-Kit and Gibson, Quinn D. and Zhou, Jiadong and Liu, Zheng and Watanabe, Kenji and Taniguchi, Takashi and Lin, Hsin and Cava, Robert J. and Fu, Liang and Gedik, Nuh and Jarillo-Herrero, Pablo},
   title = {Electrically switchable Berry curvature dipole in the monolayer topological insulator WTe2},
   journal = {Nature Physics},
   volume = {14},
   number = {9},
   pages = {900-906},
   ISSN = {1745-2481},
   DOI = {10.1038/s41567-018-0189-6},
   url = {https://doi.org/10.1038/s41567-018-0189-6},
   year = {2018},
   type = {Journal Article}
}

@Article{Godinho2018,
	author={Godinho, J.
		and Reichlov{\'a}, H.
		and Kriegner, D.
		and Nov{\'a}k, V.
		and Olejn{\'i}k, K.
		and Ka{\v{s}}par, Z.
		and {\v{S}}ob{\'a}{\v{n}}, Z.
		and Wadley, P.
		and Campion, R. P.
		and Otxoa, R. M.
		and Roy, P. E.
		and {\v{Z}}elezn{\'y}, J.
		and Jungwirth, T.
		and Wunderlich, J.},
	title={Electrically induced and detected N{\'e}el vector reversal in a collinear antiferromagnet},
	journal={Nature Communications},
	year={2018},
	month={Nov},
	day={08},
	volume={9},
	number={1},
	pages={4686},
	issn={2041-1723},
	doi={10.1038/s41467-018-07092-2},
	url={https://doi.org/10.1038/s41467-018-07092-2}
}

@article{Mehraeen2024,
  author = {Mehraeen, M.},
  title = {Quantum kinetic theory of quadratic responses},
  journal = {Phys. Rev. B},
  volume = {110},
  pages = {174423},
  year = {2024},
  doi = {10.1103/PhysRevB.110.174423}
}

@article{Ahmad2026,
  title = {Length--velocity gauge equivalence of quantum geometric nonlinear conductivity in the adiabatic DC limit},
  author = {Ahmad, Shakeel and Xue, Fei},
  journal = {Phys. Rev. B},
  volume = {114},
  issue = {15},
  pages = {154428},
  numpages = {22},
  year = {2026},
  month = {Sep},
  publisher = {American Physical Society},
  doi = {10.1103/r23j-t8kt},
  url = {https://link.aps.org/doi/10.1103/r23j-t8kt}
}

@article{Ahn2022,
   author = {Ahn, Junyeong and Guo, Guang-Yu and Nagaosa, Naoto and Vishwanath, Ashvin},
   title = {Riemannian geometry of resonant optical responses},
   journal = {Nature Physics},
   volume = {18},
   number = {3},
   pages = {290-295},
   ISSN = {1745-2481},
   DOI = {10.1038/s41567-021-01465-z},
   url = {https://doi.org/10.1038/s41567-021-01465-z},
   year = {2022},
   type = {Journal Article}
}

@article{Souza2018,
  title = {Ab initio calculation of the shift photocurrent by Wannier interpolation},
  author = {Iba\~nez-Azpiroz, Julen and Tsirkin, Stepan S. and Souza, Ivo},
  journal = {Phys. Rev. B},
  volume = {97},
  issue = {24},
  pages = {245143},
  numpages = {13},
  year = {2018},
  month = {Jun},
  publisher = {American Physical Society},
  doi = {10.1103/PhysRevB.97.245143},
  url = {https://link.aps.org/doi/10.1103/PhysRevB.97.245143}
}

@article{Cheng2024,
  title = {Lossless Spin-Orbit Torque in Antiferromagnetic Topological Insulator ${\mathrm{MnBi}}_{2}{\mathrm{Te}}_{4}$},
  author = {Tang, Junyu and Cheng, Ran},
  journal = {Phys. Rev. Lett.},
  volume = {132},
  issue = {13},
  pages = {136701},
  numpages = {7},
  year = {2024},
  month = {Mar},
  publisher = {American Physical Society},
  doi = {10.1103/PhysRevLett.132.136701},
  url = {https://link.aps.org/doi/10.1103/PhysRevLett.132.136701}
}

@article{Xu2019,
author = {Jiaheng Li  and Yang Li  and Shiqiao Du  and Zun Wang  and Bing-Lin Gu  and Shou-Cheng Zhang  and Ke He  and Wenhui Duan  and Yong Xu },
title = {Intrinsic magnetic topological insulators in van der Waals layered MnBi$_2$Te$_4$-family materials},
journal = {Science Advances},
volume = {5},
number = {6},
pages = {eaaw5685},
year = {2019},
doi = {10.1126/sciadv.aaw5685},
URL = {https://www.science.org/doi/abs/10.1126/sciadv.aaw5685},
eprint = {https://www.science.org/doi/pdf/10.1126/sciadv.aaw5685}}

@article{Mei2024,
  title = {Electrical control of intrinsic nonlinear Hall effect in antiferromagnetic topological insulator sandwiches},
  author = {Mei, Ruobing and Kaplan, Daniel and Yan, Binghai and Chang, Cui-Zu and Liu, Chao-Xing},
  journal = {Phys. Rev. B},
  volume = {110},
  issue = {16},
  pages = {165401},
  numpages = {7},
  year = {2024},
  month = {Oct},
  publisher = {American Physical Society},
  doi = {10.1103/PhysRevB.110.165401},
  url = {https://link.aps.org/doi/10.1103/PhysRevB.110.165401}
}

@article{Prodan2009,
  title = {Robustness of the spin-Chern number},
  author = {Prodan, Emil},
  journal = {Phys. Rev. B},
  volume = {80},
  issue = {12},
  pages = {125327},
  numpages = {7},
  year = {2009},
  month = {Sep},
  publisher = {American Physical Society},
  doi = {10.1103/PhysRevB.80.125327},
  url = {https://link.aps.org/doi/10.1103/PhysRevB.80.125327}
}

@article{Prodan2011,
doi = {10.1088/1751-8113/44/11/113001},
url = {https://doi.org/10.1088/1751-8113/44/11/113001},
year = {2011},
month = {feb},
publisher = {},
volume = {44},
number = {11},
pages = {113001},
author = {Prodan, Emil},
title = {Disordered topological insulators: a non-commutative geometry perspective},
journal = {Journal of Physics A: Mathematical and Theoretical}
}

@article{Review_QuantumGeometry,
  title = {Quantum geometry phenomena in condensed matter systems},
  author = {Gao, Anyuan and Nagaosa, Naoto and Ni, Ni and Xu, Su-Yang},
  journal = {Rev. Mod. Phys.},
  volume = {98},
  issue = {3},
  pages = {035005},
  numpages = {67},
  year = {2026},
  month = {Sep},
  publisher = {American Physical Society},
  doi = {10.1103/f78t-ky69},
  url = {https://link.aps.org/doi/10.1103/f78t-ky69}
}

@article{He2026,
  title = {Second harmonic Hall response in insulators: Interband quantum geometry and breakdown of Kleinman's conjecture},
  author = {He, Wen-Yu and Law, K. T.},
  journal = {Phys. Rev. B},
  volume = {113},
  issue = {4},
  pages = {L041110},
  numpages = {8},
  year = {2026},
  month = {Jan},
  publisher = {American Physical Society},
  doi = {10.1103/vkml-m2sm},
  url = {https://link.aps.org/doi/10.1103/vkml-m2sm}
}

@article{Jiang_2025,
doi = {10.1088/1361-6633/ade454},
url = {https://doi.org/10.1088/1361-6633/ade454},
year = {2025},
month = {jul},
publisher = {IOP Publishing},
volume = {88},
number = {7},
pages = {076502},
author = {Jiang, Yiyang and Holder, Tobias and Yan, Binghai},
title = {Revealing quantum geometry in nonlinear quantum materials},
journal = {Reports on Progress in Physics}
}

@article{
Bansil2024,
author = {Barun Ghosh  and Yugo Onishi  and Su-Yang Xu  and Hsin Lin  and Liang Fu  and Arun Bansil },
title = {Probing quantum geometry through optical conductivity and magnetic circular dichroism},
journal = {Science Advances},
volume = {10},
number = {51},
pages = {eado1761},
year = {2024},
doi = {10.1126/sciadv.ado1761},
URL = {https://www.science.org/doi/abs/10.1126/sciadv.ado1761},
eprint = {https://www.science.org/doi/pdf/10.1126/sciadv.ado1761}}

@article{Refael2022,
  title = {Shift-current response as a probe of quantum geometry and electron-electron interactions in twisted bilayer graphene},
  author = {Chaudhary, Swati and Lewandowski, Cyprian and Refael, Gil},
  journal = {Phys. Rev. Res.},
  volume = {4},
  issue = {1},
  pages = {013164},
  numpages = {14},
  year = {2022},
  month = {Feb},
  publisher = {American Physical Society},
  doi = {10.1103/PhysRevResearch.4.013164},
  url = {https://link.aps.org/doi/10.1103/PhysRevResearch.4.013164}
}

@article{Fiete2024,
  title = {Quantum geometry and nonlinear optical responses in rhombohedral trilayer graphene},
  author = {Postlewaite, Abigail and Raj, Arpit and Chaudhary, Swati and Fiete, Gregory A.},
  journal = {Phys. Rev. B},
  volume = {110},
  issue = {24},
  pages = {245122},
  numpages = {15},
  year = {2024},
  month = {Dec},
  publisher = {American Physical Society},
  doi = {10.1103/PhysRevB.110.245122},
  url = {https://link.aps.org/doi/10.1103/PhysRevB.110.245122}
}

@article{Wu2025,
  title = {Quantum geometry probed by chiral excitonic optical response of Chern insulators},
  author = {Qiu, Wen-Xuan and Wu, Fengcheng},
  journal = {Phys. Rev. B},
  volume = {111},
  issue = {12},
  pages = {L121104},
  numpages = {7},
  year = {2025},
  month = {Mar},
  publisher = {American Physical Society},
  doi = {10.1103/PhysRevB.111.L121104},
  url = {https://link.aps.org/doi/10.1103/PhysRevB.111.L121104}
}

@article{Claassen2026,
  title = {Probing the Quantum Geometry of Correlated Metals Using Optical Conductivity},
  author = {Carmichael, Deven P. and Claassen, Martin},
  journal = {Phys. Rev. Lett.},
  volume = {137},
  issue = {8},
  pages = {086503},
  numpages = {8},
  year = {2026},
  month = {Aug},
  publisher = {American Physical Society},
  doi = {10.1103/hrxh-y5dl},
  url = {https://link.aps.org/doi/10.1103/hrxh-y5dl}
}

@article{Gao2023,
author = {Anyuan Gao  and Yu-Fei Liu  and Jian-Xiang Qiu  and Barun Ghosh  and Thaís V. Trevisan  and Yugo Onishi  and Chaowei Hu  and Tiema Qian  and Hung-Ju Tien  and Shao-Wen Chen  and Mengqi Huang  and Damien Bérubé  and Houchen Li  and Christian Tzschaschel  and Thao Dinh  and Zhe Sun  and Sheng-Chin Ho  and Shang-Wei Lien  and Bahadur Singh  and Kenji Watanabe  and Takashi Taniguchi  and David C. Bell  and Hsin Lin  and Tay-Rong Chang  and Chunhui Rita Du  and Arun Bansil  and Liang Fu  and Ni Ni  and Peter P. Orth  and Qiong Ma  and Su-Yang Xu },
title = {Quantum metric nonlinear Hall effect in a topological antiferromagnetic heterostructure},
journal = {Science},
volume = {381},
number = {6654},
pages = {181-186},
year = {2023},
doi = {10.1126/science.adf1506},
URL = {https://www.science.org/doi/abs/10.1126/science.adf1506},
eprint = {https://www.science.org/doi/pdf/10.1126/science.adf1506}}

@article{Hsieh2021,
   author = "Orenstein, J. and Moore, J.E. and Morimoto, T. and Torchinsky, D.H. and Harter, J.W. and Hsieh, D.",
   title = "Topology and Symmetry of Quantum Materials via Nonlinear Optical Responses", 
   journal= "Annual Review of Condensed Matter Physics",
   year = "2021",
   volume = "12",
   number = "Volume 12, 2021",
   pages = "247-272",
   doi = "https://doi.org/10.1146/annurev-conmatphys-031218-013712",
   url = "https://www.annualreviews.org/content/journals/10.1146/annurev-conmatphys-031218-013712",
   publisher = "Annual Reviews",
   issn = "1947-5462",
   type = "Journal Article",
  }

@article{Song2023,
   author = {Ma, Qiong and Krishna Kumar, Roshan and Xu, Su-Yang and Koppens, Frank H. L. and Song, Justin C. W.},
   title = {Photocurrent as a multiphysics diagnostic of quantum materials},
   journal = {Nature Reviews Physics},
   volume = {5},
   number = {3},
   pages = {170-184},
   ISSN = {2522-5820},
   DOI = {10.1038/s42254-022-00551-2},
   url = {https://doi.org/10.1038/s42254-022-00551-2},
   year = {2023},
   type = {Journal Article}
}

@article{Yu2025,
   author = {Yu, Jiabin and Bernevig, B. Andrei and Queiroz, Raquel and Rossi, Enrico and Törmä, Päivi and Yang, Bohm-Jung},
   title = {Quantum geometry in quantum materials},
   journal = {npj Quantum Materials},
   volume = {10},
   number = {1},
   pages = {101},
   ISSN = {2397-4648},
   DOI = {10.1038/s41535-025-00801-3},
   url = {https://doi.org/10.1038/s41535-025-00801-3},
   year = {2025},
   type = {Journal Article}
}

@article{Nagaosa2016,
author = {Takahiro Morimoto  and Naoto Nagaosa },
title = {Topological nature of nonlinear optical effects in solids},
journal = {Science Advances},
volume = {2},
number = {5},
pages = {e1501524},
year = {2016},
doi = {10.1126/sciadv.1501524},
URL = {https://www.science.org/doi/abs/10.1126/sciadv.1501524},
eprint = {https://www.science.org/doi/pdf/10.1126/sciadv.1501524}}

@article{Holder2020,
  title = {Consequences of time-reversal-symmetry breaking in the light-matter interaction: Berry curvature, quantum metric, and diabatic motion},
  author = {Holder, Tobias and Kaplan, Daniel and Yan, Binghai},
  journal = {Phys. Rev. Res.},
  volume = {2},
  issue = {3},
  pages = {033100},
  numpages = {18},
  year = {2020},
  month = {Jul},
  publisher = {American Physical Society},
  doi = {10.1103/PhysRevResearch.2.033100},
  url = {https://link.aps.org/doi/10.1103/PhysRevResearch.2.033100}
}

@article{Moore2025,
  title = {Multistate Geometry of Shift Current and Polarization},
  author = {Avdoshkin, Alexander and Mitscherling, Johannes and Moore, Joel E.},
  journal = {Phys. Rev. Lett.},
  volume = {135},
  issue = {6},
  pages = {066901},
  numpages = {8},
  year = {2025},
  month = {Aug},
  publisher = {American Physical Society},
  doi = {10.1103/w761-8nf7},
  url = {https://link.aps.org/doi/10.1103/w761-8nf7}
}

@article{Okyay2022,
   author = {Okyay, Mahmut Sait and Sato, Shunsuke A. and Kim, Kun Woo and Yan, Binghai and Jin, Hosub and Park, Noejung},
   title = {Second harmonic Hall responses of insulators as a probe of Berry curvature dipole},
   journal = {Communications Physics},
   volume = {5},
   number = {1},
   pages = {303},
   ISSN = {2399-3650},
   DOI = {10.1038/s42005-022-01086-9},
   url = {https://doi.org/10.1038/s42005-022-01086-9},
   year = {2022},
   type = {Journal Article}
}

@article{Bhalla2022,
  title = {Resonant Second-Harmonic Generation as a Probe of Quantum Geometry},
  author = {Bhalla, Pankaj and Das, Kamal and Culcer, Dimitrie and Agarwal, Amit},
  journal = {Phys. Rev. Lett.},
  volume = {129},
  issue = {22},
  pages = {227401},
  numpages = {8},
  year = {2022},
  month = {Nov},
  publisher = {American Physical Society},
  doi = {10.1103/PhysRevLett.129.227401},
  url = {https://link.aps.org/doi/10.1103/PhysRevLett.129.227401}
}

@article{Hughes1996,
  title = {Calculation of second-order optical response in semiconductors},
  author = {Hughes, James L. P. and Sipe, J. E.},
  journal = {Phys. Rev. B},
  volume = {53},
  issue = {16},
  pages = {10751--10763},
  numpages = {0},
  year = {1996},
  month = {Apr},
  publisher = {American Physical Society},
  doi = {10.1103/PhysRevB.53.10751},
  url = {https://link.aps.org/doi/10.1103/PhysRevB.53.10751}
}

@article{Otrokov2019,
   author = {Otrokov, M. M. and Klimovskikh, I. I. and Bentmann, H. and Estyunin, D. and Zeugner, A. and Aliev, Z. S. and Gaß, S. and Wolter, A. U. B. and Koroleva, A. V. and Shikin, A. M. and Blanco-Rey, M. and Hoffmann, M. and Rusinov, I. P. and Vyazovskaya, A. Yu and Eremeev, S. V. and Koroteev, Yu M. and Kuznetsov, V. M. and Freyse, F. and Sánchez-Barriga, J. and Amiraslanov, I. R. and Babanly, M. B. and Mamedov, N. T. and Abdullayev, N. A. and Zverev, V. N. and Alfonsov, A. and Kataev, V. and Büchner, B. and Schwier, E. F. and Kumar, S. and Kimura, A. and Petaccia, L. and Di Santo, G. and Vidal, R. C. and Schatz, S. and Kißner, K. and Ünzelmann, M. and Min, C. H. and Moser, Simon and Peixoto, T. R. F. and Reinert, F. and Ernst, A. and Echenique, P. M. and Isaeva, A. and Chulkov, E. V.},
   title = {Prediction and observation of an antiferromagnetic topological insulator},
   journal = {Nature},
   volume = {576},
   number = {7787},
   pages = {416-422},
   ISSN = {1476-4687},
   DOI = {10.1038/s41586-019-1840-9},
   url = {https://doi.org/10.1038/s41586-019-1840-9},
   year = {2019},
   type = {Journal Article}
}

@article{Gao2021,
   author = {Gao, Anyuan and Liu, Yu-Fei and Hu, Chaowei and Qiu, Jian-Xiang and Tzschaschel, Christian and Ghosh, Barun and Ho, Sheng-Chin and Bérubé, Damien and Chen, Rui and Sun, Haipeng and Zhang, Zhaowei and Zhang, Xin-Yue and Wang, Yu-Xuan and Wang, Naizhou and Huang, Zumeng and Felser, Claudia and Agarwal, Amit and Ding, Thomas and Tien, Hung-Ju and Akey, Austin and Gardener, Jules and Singh, Bahadur and Watanabe, Kenji and Taniguchi, Takashi and Burch, Kenneth S. and Bell, David C. and Zhou, Brian B. and Gao, Weibo and Lu, Hai-Zhou and Bansil, Arun and Lin, Hsin and Chang, Tay-Rong and Fu, Liang and Ma, Qiong and Ni, Ni and Xu, Su-Yang},
   title = {Layer Hall effect in a 2D topological axion antiferromagnet},
   journal = {Nature},
   volume = {595},
   number = {7868},
   pages = {521-525},
   ISSN = {1476-4687},
   DOI = {10.1038/s41586-021-03679-w},
   url = {https://doi.org/10.1038/s41586-021-03679-w},
   year = {2021},
   type = {Journal Article}
}

@article{Hetenyi2023,
  title = {Fluctuations, uncertainty relations, and the geometry of quantum state manifolds},
  author = {Het{\'e}nyi, Bal{\'a}zs and L{\'e}vay, P{\'e}ter},
  journal = {Phys. Rev. A},
  volume = {108},
  issue = {3},
  pages = {032218},
  numpages = {13},
  year = {2023},
  month = {Sep},
  publisher = {American Physical Society},
  doi = {10.1103/PhysRevA.108.032218},
  url = {https://link.aps.org/doi/10.1103/PhysRevA.108.032218}
}

@article{Queiroz2026,
   author = {Verma, Nishchhal and Moll, Philip J. W. and Holder, Tobias and Queiroz, Raquel},
   title = {Quantum geometry and the hidden scales in materials},
   journal = {Nature Reviews Physics},
   volume = {8},
   number = {4},
   pages = {226-239},
   ISSN = {2522-5820},
   DOI = {10.1038/s42254-026-00923-y},
   url = {https://doi.org/10.1038/s42254-026-00923-y},
   year = {2026},
   type = {Journal Article}
}

@article{Xie2025,
    author = {Liu, Tianyu and Qiang, Xiao-Bin and Lu, Hai-Zhou and Xie, X C},
    title = {Quantum geometry in condensed matter},
    journal = {National Science Review},
    volume = {12},
    number = {3},
    pages = {nwae334},
    year = {2025},
    month = {03},
    issn = {2095-5138},
    doi = {10.1093/nsr/nwae334},
    url = {https://doi.org/10.1093/nsr/nwae334},
    eprint = {https://academic.oup.com/nsr/article-pdf/12/3/nwae334/59204701/nwae334.pdf},
}

@article{Rashkeev1998,
  title = {Efficient ab initio method for the calculation of frequency-dependent second-order optical response in semiconductors},
  author = {Rashkeev, Sergey N. and Lambrecht, Walter R. L. and Segall, Benjamin},
  journal = {Phys. Rev. B},
  volume = {57},
  issue = {7},
  pages = {3905--3919},
  numpages = {0},
  year = {1998},
  month = {Feb},
  publisher = {American Physical Society},
  doi = {10.1103/PhysRevB.57.3905},
  url = {https://link.aps.org/doi/10.1103/PhysRevB.57.3905}
}

@article{Prasad2024,
  title = {Magnetism-induced second-order nonlinear optical responses in multiferroic ${\mathrm{BiFeO}}_{3}$},
  author = {Prasad, Babu Baijnath and Liu, Guan-Fu and Guo, Guang-Yu},
  journal = {Phys. Rev. B},
  volume = {110},
  issue = {1},
  pages = {014422},
  numpages = {12},
  year = {2024},
  month = {Jul},
  publisher = {American Physical Society},
  doi = {10.1103/PhysRevB.110.014422},
  url = {https://link.aps.org/doi/10.1103/PhysRevB.110.014422}
}

@article{Wang2025,
doi = {10.1088/0256-307X/42/4/047302},
url = {https://doi.org/10.1088/0256-307X/42/4/047302},
year = {2025},
month = {apr},
publisher = {Chinese Physical Society and IOP Publishing Ltd},
volume = {42},
number = {4},
pages = {047302},
author = {Wang, En and Wang, Yaxian and Meng, Sheng},
title = {Quantum Metric, Topology and Second Harmonic Generation},
journal = {Chinese Physics Letters}
}

@article{Fei2020,
  title = {Giant photogalvanic effect and second-harmonic generation in magnetic axion insulators},
  author = {Fei, Ruixiang and Song, Wenshen and Yang, Li},
  journal = {Phys. Rev. B},
  volume = {102},
  issue = {3},
  pages = {035440},
  numpages = {8},
  year = {2020},
  month = {Jul},
  publisher = {American Physical Society},
  doi = {10.1103/PhysRevB.102.035440},
  url = {https://link.aps.org/doi/10.1103/PhysRevB.102.035440}
}

@article{Froese2026,
  title = {Probing quantum geometry with two-dimensional nonlinear optical spectroscopy},
  author = {Froese, Paul and Hirsbrunner, Mark R. and Kim, Yong Baek},
  journal = {Phys. Rev. B},
  volume = {113},
  issue = {20},
  pages = {205138},
  numpages = {17},
  year = {2026},
  month = {May},
  publisher = {American Physical Society},
  doi = {10.1103/t48g-4678},
  url = {https://link.aps.org/doi/10.1103/t48g-4678}
}

@article{Han2025,
  title = {Reconstructing the Wave Function of Magnetic Topological Insulators ${\mathrm{MnBi}}_{2}{\mathrm{Te}}_{4}$ and ${\mathrm{MnBi}}_{4}{\mathrm{Te}}_{7}$ Using Spin-Resolved Photoemission},
  author = {Han, Xue and Qu, Jason and Tan, Hengxin and Tao, Zicheng and Meyer, Noah M. and Kirchmann, Patrick S. and Guo, Yanfeng and Yan, Binghai and Shen, Zhi-Xun and Sobota, Jonathan A.},
  journal = {Phys. Rev. X},
  volume = {15},
  issue = {3},
  pages = {031022},
  numpages = {13},
  year = {2025},
  month = {Jul},
  publisher = {American Physical Society},
  doi = {10.1103/lfxc-lsc1},
  url = {https://link.aps.org/doi/10.1103/lfxc-lsc1}
}

@article{Gong2019,
doi = {10.1088/0256-307X/36/7/076801},
url = {https://doi.org/10.1088/0256-307X/36/7/076801},
year = {2019},
month = {jun},
publisher = {Chinese Physical Society and IOP Publishing Ltd},
volume = {36},
number = {7},
pages = {076801},
author = {Gong, Yan and Guo, Jingwen and Li, Jiaheng and Zhu, Kejing and Liao, Menghan and Liu, Xiaozhi and Zhang, Qinghua and Gu, Lin and Tang, Lin and Feng, Xiao and Zhang, Ding and Li, Wei and Song, Canli and Wang, Lili and Yu, Pu and Chen, Xi and Wang, Yayu and Yao, Hong and Duan, Wenhui and Xu, Yong and Zhang, Shou-Cheng and Ma, Xucun and Xue, Qi-Kun and He, Ke},
title = {Experimental Realization of an Intrinsic Magnetic Topological Insulator*},
journal = {Chinese Physics Letters}
}
\end{document}


\title{Quantum-Geometric Subgap Second-Harmonic Generation
in $\mathcal{PT}$-Symmetric Antiferromagnetic Insulators: Supplemental Materials}

\author{Shakeel Ahmad\orcidlink{0009-0000-4994-9717}}
\affiliation{Department of Physics, University of Alabama at Birmingham, Birmingham 35294, AL, USA}

\author{Fei Xue\,\orcidlink{0000-0002-1737-2332}} 
\affiliation{Department of Physics, University of Alabama at Birmingham, Birmingham 35294, AL, USA}
\date{\today}

\maketitle
\section{Derivation of the second-harmonic conductivity}
\label{sec:SHG_derivation}

We derive the second-harmonic generation (SHG) conductivity in
Eq.~(2) of the main text using the
length-gauge density-matrix formulation of Ref.~\cite{Ahmad2026}.
Starting from the second-order density matrix, we specialize to a zero-temperature insulator, recast the resulting current in a gauge-invariant form, and establish its equivalence to the conventional SHG susceptibility.

\textit{Definitions and response convention.}---
The Hamiltonian and current density are
\begin{equation}
H(t)=H_0+H_E(t),\qquad H_E(t)=-qE^a(t)x^a,
\qquad j^a(t)=q\,\mathrm{Tr}[v^a\rho(t)],
\label{eq:H_total}
\end{equation}
where $q=-e$ with $e>0$, and repeated Cartesian indices are summed.
Since $[x^a,x^b]=0$, the velocity operator is independent of the electric field, $v^a=(i/\hbar)[H_0,x^a]$.

For the Bloch eigenstates $H_0(\mathbf{k})|u_n(\mathbf{k})\rangle
=\varepsilon_n(\mathbf{k})|u_n(\mathbf{k})\rangle$,
we define
\begin{equation}
\omega_{mn}=\frac{\varepsilon_m-\varepsilon_n}{\hbar},\qquad
f_{mn}=f_m-f_n,\qquad
\Delta^a_{mn}=\partial_a\omega_{mn}=v^a_{mm}-v^a_{nn},
\qquad \partial_a=\frac{\partial}{\partial k_a}.
\label{eq:SHG_band_definitions}
\end{equation}
The Berry connection, interband position matrix, and generalized
derivative are defined consistently with the main text:
\begin{equation}
\mathcal{A}^a_{mn}=i\langle u_m|\partial_a u_n\rangle,\qquad
r^a_{mn}=(1-\delta_{mn})\mathcal{A}^a_{mn},\qquad
r^{a;b}_{mn}=\partial_b r^a_{mn}
-i(\mathcal{A}^b_{mm}-\mathcal{A}^b_{nn})r^a_{mn}.
\label{eq:SHG_connection_definitions}
\end{equation}
These definitions give
$v^a_{nn}=\partial_a\varepsilon_n/\hbar$ and
$v^a_{mn}=i\omega_{mn}r^a_{mn}$ for $m\ne n$.
With the density-matrix convention
$\rho_{mn}=\langle c_n^\dagger c_m\rangle$,
the second-order current separates into diagonal
and off-diagonal contributions,
\begin{equation}
j^{a(2)}(\Omega)
=q\sum_n v^a_{nn}\rho^{(2)}_{nn}(\Omega)
+q\sum_{m\ne n}v^a_{nm}\rho^{(2)}_{mn}(\Omega).
\label{eq:2ndcurrent}
\end{equation}
For the two-dimensional system considered here,
an overall momentum integral
$\int_{\mathrm{BZ}}d^2k/(2\pi)^2$
is implicit in the current. Intermediate-band sums
are evaluated at the same momentum, without additional
momentum integrations. Spin is included in the band
indices, so no extra spin multiplicity is introduced.

The formulas are written in nondegenerate-band notation
for compactness. For $\mathcal{PT}$-protected degeneracies,
the gauge-invariant response is understood in terms of
distinct energy subspaces, with matrix-valued inter-subspace
connections, non-Abelian covariant derivatives, and
traces over closed products, as detailed below.

We adopt the Fourier convention
$E^a(t)=E^a(\omega)e^{-i\omega t}+\mathrm{c.c.}$
and define the SHG conductivity by
\begin{equation}
j^{a(2)}(2\omega)
=\sigma^{abc}(2\omega;\omega,\omega)
E^b(\omega)E^c(\omega).
\end{equation}
In the intermediate expressions, we retain two
incident frequencies $\omega_1$ and $\omega_2$,
with $\Omega=\omega_1+\omega_2$.
Retarded denominators are understood, with
$\omega_j\to\omega_j+i0^+$, so that the output
frequency in SHG is continued as $2\omega+2i0^+$.

\textit{Second-order density matrix.}---
Following Ref.~\cite{Ahmad2026}, the diagonal
and off-diagonal density-matrix elements are
\begin{equation}
\begin{aligned}
\rho^{(2)}_{nn}(\Omega;\omega_1,\omega_2)
={}&\frac{q^2E^b(\omega_1)E^c(\omega_2)}{\hbar^2\Omega}
\Bigg[-\frac{\partial_b\partial_c f_n}{\omega_2}
+\sum_{\ell\ne n}f_{\ell n}
\left(\frac{r^b_{n\ell}r^c_{\ell n}}{\omega_{\ell n}-\omega_2}
-\frac{r^c_{n\ell}r^b_{\ell n}}{\omega_{\ell n}+\omega_2}\right)
+(b,\omega_1)\leftrightarrow(c,\omega_2)\Bigg],
\end{aligned}
\label{eq:diagonaldensity}
\end{equation}
\begin{equation}
\rho^{(2)}_{mn}(\Omega;\omega_1,\omega_2)
=-\frac{q^2E^b(\omega_1)E^c(\omega_2)}{\hbar^2(\omega_{mn}-\Omega)}
\left[\mathcal K^{bc}_{mn}(\omega_2)
+(b,\omega_1)\leftrightarrow(c,\omega_2)\right],\qquad m\ne n,
\label{eq:offdiagdensity}
\end{equation}
where
\begin{equation}
\begin{aligned}
\mathcal K^{bc}_{mn}(\omega)
={}&i r^{c;b}_{mn}\frac{f_{mn}}{\omega_{mn}-\omega}
+i r^c_{mn}\partial_b\left(\frac{f_{mn}}{\omega_{mn}-\omega}\right)
-\frac{i}{\omega}r^b_{mn}\partial_c f_{mn}
\\
&+\sum_{\ell\ne m,n}
\left(\frac{f_{\ell n}r^b_{m\ell}r^c_{\ell n}}{\omega_{\ell n}-\omega}
-\frac{f_{m\ell}r^c_{m\ell}r^b_{\ell n}}{\omega_{m\ell}-\omega}\right).
\end{aligned}
\label{eq:SHG_density_kernel}
\end{equation}
In Eq.~\eqref{eq:diagonaldensity}, the exchange
$(b,\omega_1)\leftrightarrow(c,\omega_2)$
acts on the terms inside the brackets, leaving
the common electric-field prefactor unchanged.
These expressions provide the starting point
for the insulating SHG limit derived below.

\textit{SHG in an insulator.}---
We now specialize to SHG by setting
$\omega_1=\omega_2=\omega$ and $\Omega=2\omega$.
For a zero-temperature insulator with the chemical
potential in the gap, the band occupations are
independent of momentum, so $\partial_a f_n=0$
and all occupation-derivative terms vanish.
For convenience, define
\begin{equation}
F_{mn}(\omega)=\frac{f_{mn}}{\omega_{mn}-\omega},\qquad
B_{mn}(2\omega)=\frac{2\omega}{\omega_{mn}-2\omega},
\label{eq:FB_def_three_band}
\end{equation}
and the primed commutator
\begin{equation}
[r^b,Fr^c]'_{mn}
=\sum_{\ell\ne m,n}
\left(r^b_{m\ell}F_{\ell n}r^c_{\ell n}
-F_{m\ell}r^c_{m\ell}r^b_{\ell n}\right).
\label{eq:SHG_primed_commutator}
\end{equation}
Hereafter, the frequency arguments of $F_{mn}$
and $B_{mn}$ are suppressed unless needed.

Adding the two incident-field orderings combines the
resonance denominators and removes the apparent
$1/(2\omega)$ singularity in the diagonal density
matrix. Exchanging the band indices in one of the
resulting terms then replaces the group velocity by
the velocity difference,
\begin{equation}
\begin{aligned}
\sigma_{\rm diag}^{abc}
&=\frac{q^3}{\hbar^2}\sum_{m\ne n}
\frac{f_{mn}v^a_{nn}
(r^b_{nm}r^c_{mn}+r^c_{nm}r^b_{mn})}
{\omega_{mn}^2-\omega^2}
\\
&=\frac{q^3}{\hbar^2}\sum_{m\ne n}
\frac{f_{mn}\Delta^a_{nm}r^b_{nm}r^c_{mn}}
{\omega_{mn}^2-\omega^2}.
\end{aligned}
\label{eq:sigma_total1}
\end{equation}

To simplify the off-diagonal contribution,
we introduce
\begin{equation}
\Gamma^{abc}_{mn}
=r^a_{nm}r^{c;b}_{mn}F_{mn}
+r^a_{nm}r^c_{mn}\partial_bF_{mn}
-i r^a_{nm}[r^b,Fr^c]'_{mn}+(b\leftrightarrow c).
\label{eq:gamma-func}
\end{equation}
Using $v^a_{nm}=-i\omega_{mn}r^a_{nm}$ in
Eq.~\eqref{eq:2ndcurrent}, together with
$\omega_{mn}/(\omega_{mn}-2\omega)=1+B_{mn}$ gives
\begin{equation}
\sigma^{abc}
=\frac{q^3}{\hbar^2}\sum_{m\ne n}
\left[\frac{f_{mn}\Delta^a_{nm}r^b_{nm}r^c_{mn}}
{\omega_{mn}^2-\omega^2}
-\Gamma^{abc}_{mn}-B_{mn}\Gamma^{abc}_{mn}\right].
\label{eq:sigma_total_02}
\end{equation}

We next reduce the two terms involving $\Gamma^{abc}_{mn}$
using integration by parts over the Brillouin zone.
The required identities are
\begin{equation}
\partial_b(r^a_{nm}r^c_{mn})
=r^{a;b}_{nm}r^c_{mn}+r^a_{nm}r^{c;b}_{mn},\qquad
r^{b;a}_{nm}-r^{a;b}_{nm}
=i\sum_{\ell\ne m,n}
(r^a_{n\ell}r^b_{\ell m}-r^b_{n\ell}r^a_{\ell m}).
\label{eq:SHG_derivative_identities}
\end{equation}
Relabeling the intermediate-band sums gives the useful identity
\begin{equation}
\sum_{m\ne n}i r^a_{nm}[r^b,Fr^c]'_{mn}
=\sum_{m\ne n}F_{mn}
(r^{b;a}_{nm}-r^{a;b}_{nm})r^c_{mn}.
\label{eq:SHG_commutator_relabeling}
\end{equation}
Combining these identities and integrating by parts yields
\begin{equation}
\sum_{m\ne n}\Gamma^{abc}_{mn}
=-\sum_{m\ne n}F_{mn}
\left(r^{b;a}_{nm}r^c_{mn}+r^{c;a}_{nm}r^b_{mn}\right).
\label{eq:SHG_gamma_reduction}
\end{equation}
This term is generally nonzero and supplies the first
generalized-derivative term in the final conductivity.
Similarly, the $B_{mn}$-weighted contribution becomes

\begin{equation}
-\sum_{m\ne n}B_{mn}\Gamma^{abc}_{mn}
=\sum_{m\ne n}\left[
F_{mn}B_{mn}r^{a;b}_{nm}r^c_{mn}
+F_{mn}(\partial_bB_{mn})r^a_{nm}r^c_{mn}
+iB_{mn}r^a_{nm}[r^b,Fr^c]'_{mn}
+(b\leftrightarrow c)\right].
\label{eq:SHG_Bgamma_reduction}
\end{equation}
These reductions hold after Brillouin-zone integration,
assuming periodicity or vanishing boundary terms.

Substituting Eqs.~\eqref{eq:SHG_gamma_reduction}
and~\eqref{eq:SHG_Bgamma_reduction} into
Eq.~\eqref{eq:sigma_total_02}, we obtain
\begin{equation}
\begin{aligned}
\sigma^{abc}(2\omega;\omega,\omega)
=\frac{q^3}{\hbar^2}\sum_{m\ne n}\Bigg\{
&\frac{f_{mn}\Delta^a_{nm}r^b_{nm}r^c_{mn}}
{\omega_{mn}^2-\omega^2}
+\Bigg[F_{mn}r^{b;a}_{nm}r^c_{mn}
+B_{mn}F_{mn}r^{a;b}_{nm}r^c_{mn}
\\
&+F_{mn}(\partial_bB_{mn})r^a_{nm}r^c_{mn}
+iB_{mn}r^a_{nm}[r^b,Fr^c]'_{mn}
+(b\leftrightarrow c)\Bigg]\Bigg\},
\end{aligned}
\label{eq:SI_final_sigma_total}
\end{equation}
Here
$\partial_bB_{mn}
=-2\omega\Delta^b_{mn}/(\omega_{mn}-2\omega)^2$,
and $(b\leftrightarrow c)$ acts only within the
square brackets. Each contribution is gauge invariant,
with closed products understood as traces in degenerate
subspaces. The primed commutator describes virtual
processes involving three distinct energy subspaces
and vanishes in a two-band model.
For the fully subgap response, we impose
$2\hbar\omega<E_{\rm g}$, where $E_{\rm g}$ is the band gap.
In this clean, off-resonant regime, SHG arises entirely
from virtual interband processes.

In the static limit, band exchange and the
generalized-derivative product rule reduce the
conductivity to a total derivative,
\begin{equation}
\sigma^{abc}(0;0,0)
=\frac{q^3}{\hbar^2}\sum_{m\ne n}
\partial_a\left(\frac{f_{mn}r^b_{nm}r^c_{mn}}{\omega_{mn}}\right)
=0,
\label{eq:SHG_static_limit}
\end{equation}
where the final equality follows from Brillouin-zone
periodicity, or vanishing boundary contributions.
This agrees with the insulating dc limit of
Ref.~\cite{Ahmad2026}.

\textit{Degenerate energy subspaces.}---
For exactly degenerate multiplets, including
$\mathcal{PT}$-related pairs, the band indices in
the preceding expressions label distinct energy
subspaces. The inter-subspace Berry connections
are matrices, and their generalized derivatives
take the non-Abelian covariant form
\begin{equation}
r^{a;b}_{mn}
=\partial_b r^a_{mn}
-i\left(
\mathcal A^b_{mm}r^a_{mn}
-r^a_{mn}\mathcal A^b_{nn}
\right).
\label{eq:SHG_block_derivative}
\end{equation}
Each closed product is traced over its initial
energy subspace, while $m\ne n$ and
$\ell\ne m,n$ refer to distinct subspaces.
The reductions above extend to this setting
through covariant product rules and cyclicity
of the trace, ensuring gauge invariance under
unitary rotations within degenerate multiplets.

\subsection{Algebraic comparison with the conventional SHG susceptibility}
\label{sec:SHG_comparison}

To compare our conductivity with the conventional SHG
susceptibility, we define $P^a(2\omega)=\chi_{\rm ours}^{abc}E^b(\omega)E^c(\omega)$. 
With our Fourier convention,
\begin{equation}
j^a(2\omega)=-2i\omega P^a(2\omega),
\qquad
\chi_{\rm ours}^{abc}
=\frac{\sigma^{abc}(2\omega;\omega,\omega)}{-2i\omega}.
\label{eq:sigma_chi_relation}
\end{equation}

We compare the three contributions to
Eq.~\eqref{eq:SI_final_sigma_total} with the
length-gauge formulation of Sipe and
Shkrebtii~\cite{Sipe2000}, retaining the expressions
before imposing time-reversal symmetry.
We also establish their equivalence to the
frequency-dependent formulas of
Rashkeev et al.~\cite{Rashkeev1998} and their
magnetic extension by Prasad
et al.~\cite{Prasad2024}.
The overall normalization is discussed below.
In the literature notation
$\chi^{abc}(-2\omega;\omega,\omega)$,
the argument $-2\omega$ labels the output-frequency
slot, while the physical output frequency is $2\omega$.

\textit{Three-band contribution.}---
Expanding the primed commutator in
Eq.~\eqref{eq:SI_final_sigma_total} gives
\begin{equation}
\frac{\sigma^{abc}_{3{\rm b}}}{-2i\omega}
=\frac{q^3}{\hbar^2}
\sum_{\substack{m\ne n\\\ell\ne m,n}}
\frac{r^a_{nm}}{\omega_{mn}-2\omega}
\left(\frac{f_{m\ell}}{\omega_{m\ell}-\omega}
-\frac{f_{\ell n}}{\omega_{\ell n}-\omega}\right)
\left(r^b_{m\ell}r^c_{\ell n}+r^c_{m\ell}r^b_{\ell n}\right).
\label{eq:chi_3b_ours}
\end{equation}

In our charge convention, the first-order
density-matrix coefficient of
Ref.~\cite{Sipe2000} is
\begin{equation}
\mathcal B^b_{mn}
=\frac{qf_{nm}r^b_{mn}}{\hbar(\omega_{mn}-\omega)}
=-\frac{q}{\hbar}F_{mn}r^b_{mn},
\label{eq:sipe_B_def}
\end{equation}
which should not be confused with the scalar
$B_{mn}$ defined in Eq.~\eqref{eq:FB_def_three_band}.
For one incident-field ordering,
their Eq.~(B1) becomes
\begin{equation}
\begin{aligned}
\chi_{2{\rm ter},(bc)}^{abc}
&=\frac{q^2}{\hbar}
\sum_{\substack{m\ne n\\\ell\ne m,n}}
\frac{r^a_{nm}}{\omega_{mn}-2\omega}
\left(
r^c_{m\ell}\mathcal B^b_{\ell n}
-\mathcal B^b_{m\ell}r^c_{\ell n}
\right)
\\
&=\frac{q^3}{\hbar^2}
\sum_{\substack{m\ne n\\\ell\ne m,n}}
\frac{r^a_{nm}}{\omega_{mn}-2\omega}
\left(
\frac{f_{m\ell}r^b_{m\ell}r^c_{\ell n}}
{\omega_{m\ell}-\omega}
-\frac{f_{\ell n}r^c_{m\ell}r^b_{\ell n}}
{\omega_{\ell n}-\omega}
\right).
\end{aligned}
\label{eq:sipe_B1}
\end{equation}
Adding the exchanged ordering $b\leftrightarrow c$
reproduces Eq.~\eqref{eq:chi_3b_ours} exactly,
without requiring Brillouin-zone integration
or time-reversal symmetry.

The same contribution can be rearranged into
the purely interband form of
Refs.~\cite{Rashkeev1998,Prasad2024}.
Using
$\omega_{mn}=\omega_{m\ell}+\omega_{\ell n}$
and $f_{m\ell}+f_{\ell n}=-f_{nm}$,
we obtain
\begin{equation}
\begin{aligned}
\frac{1}{\omega_{mn}-2\omega}
\left(
\frac{f_{m\ell}}{\omega_{m\ell}-\omega}
-\frac{f_{\ell n}}{\omega_{\ell n}-\omega}
\right)
=
\frac{1}{\omega_{\ell n}-\omega_{m\ell}}
\left(
\frac{2f_{nm}}{\omega_{mn}-2\omega}
+\frac{f_{\ell n}}{\omega_{\ell n}-\omega}
+\frac{f_{m\ell}}{\omega_{m\ell}-\omega}
\right).
\end{aligned}
\label{eq:three_band_identity}
\end{equation}
This reproduces the three-band frequency kernel
in Eq.~(2) of Ref.~\cite{Prasad2024}.

\textit{Mixed interband--intraband contribution.}---
The terms proportional to $B_{mn}F_{mn}$ and
$F_{mn}\partial_b B_{mn}$ in
Eq.~\eqref{eq:SI_final_sigma_total} define
$\sigma_{\rm pol}^{abc}$.
Using the definitions of $F_{mn}$ and $B_{mn}$,
we obtain
\begin{equation}
\frac{\sigma_{\rm pol}^{abc}}{-2i\omega}
=\frac{iq^3}{\hbar^2}\sum_{m\ne n}F_{mn}
\left[\frac{r^{a;b}_{nm}r^c_{mn}}{\omega_{mn}-2\omega}
-\frac{\Delta^b_{mn}r^a_{nm}r^c_{mn}}{(\omega_{mn}-2\omega)^2}
+(b\leftrightarrow c)\right].
\label{eq:chi_mixed_ours}
\end{equation}
For one field ordering, Eqs.~(B2) and (B3)
of Ref.~\cite{Sipe2000} give
\begin{equation}
\begin{aligned}
\widetilde\chi_{2{\rm tra},(bc)}^{abc}
&=\frac{iq^2}{\hbar}\sum_{m\ne n}
\frac{r^a_{nm}\mathcal B^{b;c}_{mn}}{\omega_{mn}-2\omega}
\\
&=\frac{iq^3}{\hbar^2}\sum_{m\ne n}f_{nm}
\left[\frac{r^a_{nm}r^{b;c}_{mn}}
{(\omega_{mn}-2\omega)(\omega_{mn}-\omega)}
-\frac{\Delta^c_{mn}r^a_{nm}r^b_{mn}}
{(\omega_{mn}-2\omega)(\omega_{mn}-\omega)^2}\right].
\end{aligned}
\label{eq:sipe_B2}
\end{equation}
Using $\partial_c f_{nm}=0$ and the generalized-derivative
product rule, the difference
between the two expressions reduces to
\begin{equation}
\begin{aligned}
\left(\widetilde\chi_{2{\rm tra},(bc)}^{abc}
+\widetilde\chi_{2{\rm tra},(cb)}^{abc}\right)
-\frac{\sigma_{\rm pol}^{abc}}{-2i\omega}
=\frac{iq^3}{\hbar^2}\sum_{m\ne n}
\left[\partial_c\left(
\frac{f_{nm}r^a_{nm}r^b_{mn}}
{(\omega_{mn}-2\omega)(\omega_{mn}-\omega)}\right)
+(b\leftrightarrow c)\right]=0.
\end{aligned}
\label{eq:sipe_B3_total_derivative}
\end{equation}
The last equality follows after Brillouin-zone
integration, assuming periodicity or vanishing
boundary terms. Thus, the two formulations
give the same integrated response, although
their momentum-resolved integrands differ.
Partial-fraction rearrangements and removal
of total derivatives recover the corresponding
mixed interband-intraband structures of
Refs.~\cite{Rashkeev1998,Prasad2024}.

\textit{Remaining current and diagonal contribution.}---
The remaining terms in our conductivity are
\begin{equation}
\sigma_{\rm rem}^{abc}
=\frac{q^3}{\hbar^2}\sum_{m\ne n}
\left[\frac{f_{mn}\Delta^a_{nm}r^b_{nm}r^c_{mn}}
{\omega_{mn}^2-\omega^2}
+F_{mn}\left(r^{b;a}_{nm}r^c_{mn}
+r^{c;a}_{nm}r^b_{mn}\right)\right].
\label{eq:SHG_check_remaining_sigma}
\end{equation}
These correspond to the intraband-current
contributions I and II in Appendix B of
Ref.~\cite{Sipe2000}, before the frequency
symmetrization leading to their Eqs.~(37)--(40).
In our notation, the ordered contributions are
\begin{equation}
\sigma_{{\rm I},(bc)}^{abc}
=-\frac{q^2}{\hbar}\sum_{m\ne n}\mathcal B^b_{mn}r^{c;a}_{nm}
=\frac{q^3}{\hbar^2}\sum_{m\ne n}F_{mn}r^{c;a}_{nm}r^b_{mn},
\label{eq:sipe_current_I_ordered}
\end{equation}
\begin{equation}
\sigma_{{\rm II},(bc)}^{abc}
=\frac{q^2}{2\hbar\omega}\sum_{m\ne n}
\Delta^a_{mn}r^c_{nm}\mathcal B^b_{mn}
=\frac{q^3}{2\hbar^2\omega}\sum_{m\ne n}
\frac{f_{nm}\Delta^a_{mn}r^c_{nm}r^b_{mn}}{\omega_{mn}-\omega}.
\label{eq:sipe_current_II_ordered}
\end{equation}
Adding the two orderings of current I reproduces
the generalized-derivative terms in
Eq.~\eqref{eq:SHG_check_remaining_sigma}.

For current II, the combination
$f_{nm}\Delta^a_{mn}$ and the symmetrized
matrix-element product are invariant under
$m\leftrightarrow n$. Pairing the original
and exchanged band sums gives
\begin{equation}
\begin{aligned}
&\frac{1}{2\omega}\sum_{m\ne n}
\frac{f_{nm}\Delta^a_{mn}
(r^b_{nm}r^c_{mn}+r^c_{nm}r^b_{mn})}{\omega_{mn}-\omega}
\\
&\quad=\frac{1}{4\omega}\sum_{m\ne n}f_{nm}\Delta^a_{mn}
(r^b_{nm}r^c_{mn}+r^c_{nm}r^b_{mn})
\left(\frac{1}{\omega_{mn}-\omega}-\frac{1}{\omega_{mn}+\omega}\right)
\\
&\quad=\frac12\sum_{m\ne n}
\frac{f_{nm}\Delta^a_{mn}
(r^b_{nm}r^c_{mn}+r^c_{nm}r^b_{mn})}{\omega_{mn}^2-\omega^2}
=\sum_{m\ne n}\frac{f_{mn}\Delta^a_{nm}r^b_{nm}r^c_{mn}}
{\omega_{mn}^2-\omega^2}.
\end{aligned}
\label{eq:sipe_current_II_band_exchange}
\end{equation}
Thus, current II reproduces the diagonal
contribution in Eq.~\eqref{eq:SHG_check_remaining_sigma}.
This identity holds at each momentum after
the band summation.

For a direct comparison with the magnetic-system
formula, consider the last two terms of Eq.~(3)
in Ref.~\cite{Prasad2024}.
Using $P=\chi EE$, the same charge convention
and momentum measure, and retaining the conventional
susceptibility normalization, these terms read
\begin{equation}
\chi_{{\rm P,rem}}^{abc}
=\frac{iq^3}{2\hbar^2}\sum_{m\ne n}f_{nm}
\left[-\frac{r^{b;a}_{nm}r^c_{mn}+r^{c;a}_{nm}r^b_{mn}}
{2\omega_{mn}(\omega_{mn}-\omega)}
+\frac{\Delta^a_{mn}(r^b_{nm}r^c_{mn}+r^c_{nm}r^b_{mn})}
{4\omega_{mn}^2(\omega_{mn}-\omega)}\right].
\label{eq:Prasad_remaining_chi}
\end{equation}
After band exchange, our remaining contribution
can be written as
\begin{equation}
\frac{\sigma_{\rm rem}^{abc}}{-2i\omega}
=\frac{iq^3}{2\omega\hbar^2}\sum_{m\ne n}f_{nm}
\left[-\frac{r^{b;a}_{nm}r^c_{mn}+r^{c;a}_{nm}r^b_{mn}}
{\omega_{mn}-\omega}
+\frac{\Delta^a_{mn}(r^b_{nm}r^c_{mn}+r^c_{nm}r^b_{mn})}
{2\omega_{mn}(\omega_{mn}-\omega)}\right].
\label{eq:remaining_chi_for_subtraction}
\end{equation}
Subtracting twice Eq.~\eqref{eq:Prasad_remaining_chi}
eliminates the explicit frequency dependence
from the bracket:
\begin{equation}
\begin{aligned}
\frac{\sigma_{\rm rem}^{abc}}{-2i\omega}
-2\chi_{{\rm P,rem}}^{abc}
=\frac{iq^3}{2\omega\hbar^2}
\sum_{m\ne n}f_{nm}
\Bigg[
-\frac{r^{b;a}_{nm}r^c_{mn}
+r^{c;a}_{nm}r^b_{mn}}{\omega_{mn}}
+\frac{\Delta^a_{mn}
(r^b_{nm}r^c_{mn}+r^c_{nm}r^b_{mn})}
{2\omega_{mn}^2}
\Bigg].
\end{aligned}
\label{eq:remaining_chi_difference}
\end{equation}
Since $f_{nm}/\omega_{mn}$ is invariant under
$m\leftrightarrow n$, the generalized-derivative
terms satisfy
\begin{equation}
\begin{aligned}
\sum_{m\ne n}\frac{f_{nm}}{\omega_{mn}}
\left(
r^{b;a}_{nm}r^c_{mn}
+r^{c;a}_{nm}r^b_{mn}
\right)
=\frac12\sum_{m\ne n}
\frac{f_{nm}}{\omega_{mn}}
\partial_a
\left(
r^b_{nm}r^c_{mn}
+r^c_{nm}r^b_{mn}
\right).
\end{aligned}
\label{eq:sipe_intra_static_IBP}
\end{equation}
Using $\partial_a f_{nm}=0$ and
$\partial_a\omega_{mn}=\Delta^a_{mn}$,
the remaining difference becomes
\begin{equation}
\begin{aligned}
\frac{\sigma_{\rm rem}^{abc}}{-2i\omega}
-2\chi_{{\rm P,rem}}^{abc}
=-\frac{iq^3}{4\omega\hbar^2}
\sum_{m\ne n}
\partial_a\left[
\frac{f_{nm}
(r^b_{nm}r^c_{mn}+r^c_{nm}r^b_{mn})}
{\omega_{mn}}
\right]
=0.
\end{aligned}
\label{eq:remaining_Prasad_total_derivative}
\end{equation}
Again, the final equality follows after
Brillouin-zone integration.

\textit{Summary and normalization.}---
The three-band, mixed interband-intraband,
and remaining-current contributions thus reproduce
the conventional length-gauge SHG susceptibility
after the appropriate band summations and
Brillouin-zone integrations.
Importantly, the comparison does not assume
time-reversal symmetry and therefore includes
the magnetic contribution retained in
Ref.~\cite{Prasad2024}.
In the time-reversal-symmetric limit, the
last $\Delta^a_{mn}rr$ term in their mixed
susceptibility integrates to zero, recovering
the nonmagnetic formulation of
Ref.~\cite{Rashkeev1998}.

Our response tensors retain the explicit sum over both
incident-field orderings, whereas the conventional
susceptibility uses their average. With the same charge
convention and momentum measure,
\begin{equation}
\chi_{\rm ours}^{abc}
=\frac{\sigma^{abc}}{-2i\omega}
=2\chi_{\rm lit}^{abc},
\qquad
\chi^{(2),abc}_{\rm lit,SI}
=\frac{\chi_{\rm lit}^{abc}}{\epsilon_0}.
\label{eq:SHG_check_normalized_result}
\end{equation}
All conductivities and expansion coefficients reported
here use the summed-ordering normalization.

Finally, the two-dimensional momentum integration
used here gives sheet response coefficients.
Conversion to an effective bulk SI susceptibility
requires specifying the layer thickness in addition
to the factor of $\epsilon_0$.

\subsection{Subgap expansion in frequency}
\label{sec:subgap_expansion}

We expand Eq.~\eqref{eq:SI_final_sigma_total} about
$\omega=0$ to identify the geometric contributions at
successive orders in frequency.
For a zero-temperature insulator,
$\partial_a f_{mn}=0$.
In the clean, fully subgap regime
$2\hbar\omega<E_\mathrm{g}$, we write
\begin{equation}
\sigma^{abc}(2\omega;\omega,\omega)
=
\sum_{N\geq0}\omega^N\mathcal G_N^{abc},
\qquad
\sigma^{abc}_{[N]}=\omega^N\mathcal G_N^{abc}.
\label{eq:subgap_series}
\end{equation}
As in the preceding derivation, an overall
momentum integral is implicit in the band sums,
while intermediate-band sums are evaluated at
the same momentum. For degenerate multiplets,
the expressions are understood in terms of
distinct energy subspaces, with covariant
derivatives and traces over closed products.
The geometric expansions below are applied where
$2|\omega|<|\omega_{mn}|$ for the denominators
being expanded. Contributions involving removable
degeneracies must be combined before taking
the low-frequency limit.

The reality of the time-dependent response requires
\begin{equation}
\begin{aligned}
\sigma^{abc}(-2\omega;-\omega,-\omega)
&=
\big[\sigma^{abc}(2\omega;\omega,\omega)\big]^*,
\\
(\mathcal G_N^{abc})^*
&=(-1)^N\mathcal G_N^{abc}.
\end{aligned}
\label{eq:subgap_reality}
\end{equation}
Thus, even-order coefficients are real, whereas
odd-order coefficients are purely imaginary.
The vanishing of the latter requires additional
symmetry constraints, discussed below.

\textit{Frequency expansions and integration by parts.}---
The frequency-dependent factors in
Eq.~\eqref{eq:SI_final_sigma_total} admit the
series expansions
\begin{align}
F_{mn}(\omega)
&=
\frac{f_{mn}}{\omega_{mn}}
\sum_{k=0}^{\infty}
\left(\frac{\omega}{\omega_{mn}}\right)^k,
\label{eq:Fmn_expansion}
\\
B_{mn}(2\omega)
&=
\sum_{k=1}^{\infty}
\left(\frac{2\omega}{\omega_{mn}}\right)^k,
\label{eq:Bmn_expansion}
\\
\partial_b B_{mn}(2\omega)
&=
-\frac{\Delta^b_{mn}}{\omega_{mn}}
\sum_{k=1}^{\infty}
k\left(\frac{2\omega}{\omega_{mn}}\right)^k.
\label{eq:partial_Bmn_expansion}
\end{align}
The products entering the conductivity,
expanded through fourth order, are
\begin{align}
F_{mn}B_{mn}
&=
f_{mn}\left[
\frac{2\omega}{\omega_{mn}^2}
+
\frac{6\omega^2}{\omega_{mn}^3}
+
\frac{14\omega^3}{\omega_{mn}^4}
+
\frac{30\omega^4}{\omega_{mn}^5}
\right]
+O(\omega^5),
\label{eq:FB_product_expansion}
\\
F_{mn}\partial_b B_{mn}
&=
-f_{mn}\Delta^b_{mn}\left[
\frac{2\omega}{\omega_{mn}^3}
+
\frac{10\omega^2}{\omega_{mn}^4}
+
\frac{34\omega^3}{\omega_{mn}^5}
+
\frac{98\omega^4}{\omega_{mn}^6}
\right]
+O(\omega^5).
\label{eq:FdB_product_expansion}
\end{align}

To express the expansion coefficients in terms of
interband geometry, we eliminate the group-velocity
differences by integration by parts.
Using the interband Hermitian connection defined
in the main text,
$\mathcal C^{mn}_{abc}=r^a_{nm}r^{c;b}_{mn}$,
together with
$\partial_a\omega_{mn}^{-p}
=-p\Delta^a_{mn}\omega_{mn}^{-p-1}$,
we obtain
\begin{equation}
\begin{aligned}
&\sum_{m\ne n}
\frac{f_{mn}\Delta^a_{mn}r^b_{nm}r^c_{mn}}
{\omega_{mn}^{p+1}}
=\frac{1}{p}
\sum_{m\ne n}
\frac{f_{mn}}{\omega_{mn}^{p}}
\left(
r^{b;a}_{nm}r^c_{mn}
+r^b_{nm}r^{c;a}_{mn}
\right)
=\frac{1}{p}
\sum_{m\ne n}
\frac{f_{mn}}{\omega_{mn}^{p}}
\left(
\mathcal C^{nm}_{cab}
+\mathcal C^{mn}_{bac}
\right).
\end{aligned}
\label{eq:idenity1}
\end{equation}
A term containing
$\Delta^a_{nm}=-\Delta^a_{mn}$
has the opposite sign.
These equalities hold after Brillouin-zone
integration, assuming periodicity or vanishing
boundary terms. The two connection terms can
be brought into the same band-index ordering
using the band-exchange identity derived below.

We also introduce the fully symmetric Cartesian
part of the Hermitian connection,
\begin{equation}
\begin{aligned}
\mathcal C^{mn}_{(abc)}
\equiv\frac{1}{6}\big(
&\mathcal C^{mn}_{abc}
+\mathcal C^{mn}_{acb}
+\mathcal C^{mn}_{bac}
+\mathcal C^{mn}_{bca}
+\mathcal C^{mn}_{cab}
+\mathcal C^{mn}_{cba}
\big).
\end{aligned}
\label{eq:connection_full_symmetrization}
\end{equation}

\textit{Order-by-order coefficient extraction.}---
We separate the coefficient at each order into
a Hermitian-connection contribution and an
explicit three-band contribution,
\begin{equation}
\mathcal G_N^{abc}
=
\mathcal G_{N,\mathcal C}^{abc}
+
\mathcal G_{N,3\mathrm b}^{abc},
\qquad
\mathcal G_{0,3\mathrm b}^{abc}=0.
\label{eq:subgap_coefficient_decomposition}
\end{equation}
Here the label $\mathcal C$ denotes the terms
expressed through Hermitian connections after
integration by parts; it does not imply a
two-band approximation.

For compactness, we introduce three numerical coefficients,
\begin{align}
\alpha_N&=\frac{1+(-1)^N}{2},
\nonumber\\
\beta_N&=\sum_{s=1}^{N}2^s=2^{N+1}-2,
\nonumber\\
\gamma_N&=\sum_{s=1}^{N}s\,2^s
=(N-1)2^{N+1}+2.
\label{eq:subgap_expansion_weights}
\end{align}
Here $\alpha_N$ selects the even powers in
$(\omega_{mn}^2-\omega^2)^{-1}$, while
$\beta_N$ and $\gamma_N$ arise from
$F_{mn}B_{mn}$ and $F_{mn}\partial_bB_{mn}$,
respectively.

Collecting the coefficient of $\omega^N$ from all
terms except the explicit three-band contribution
in Eq.~\eqref{eq:SI_final_sigma_total}, we obtain
\begin{equation}
\begin{aligned}
\mathcal G_{N,\mathcal C}^{abc}
=
\frac{q^3}{\hbar^2}
\sum_{m\ne n}
\frac{f_{mn}}{\omega_{mn}^{N+1}}
\Bigg\{
&\alpha_N
\frac{\Delta^a_{nm}r^b_{nm}r^c_{mn}}
{\omega_{mn}}
+\Big[
r^{b;a}_{nm}r^c_{mn}
+\beta_N r^{a;b}_{nm}r^c_{mn}
-\gamma_N
\frac{\Delta^b_{mn}r^a_{nm}r^c_{mn}}
{\omega_{mn}}
+(b\leftrightarrow c)
\Big]\Bigg\}.
\end{aligned}
\label{eq:subgap_connection_raw_coefficient}
\end{equation}

Hermitian conjugation and band exchange give
\begin{equation}
\mathcal C^{nm}_{abc}
=
\big(\mathcal C^{mn}_{abc}\big)^*,
\qquad
\frac{f_{nm}}{\omega_{nm}^{N+1}}
=
(-1)^N\frac{f_{mn}}{\omega_{mn}^{N+1}},
\end{equation}
and therefore
\begin{equation}
\sum_{m\ne n}
\frac{f_{mn}}{\omega_{mn}^{N+1}}
\mathcal C^{nm}_{abc}
=
(-1)^N
\sum_{m\ne n}
\frac{f_{mn}}{\omega_{mn}^{N+1}}
\mathcal C^{mn}_{abc}.
\label{eq:subgap_connection_band_parity}
\end{equation}

Applying Eq.~\eqref{eq:idenity1} to the two
velocity-difference terms, followed by the
band-exchange relation
\eqref{eq:subgap_connection_band_parity}, gives
\begin{equation}
\begin{aligned}
&\sum_{m\ne n}
\frac{f_{mn}\Delta^a_{nm}
r^b_{nm}r^c_{mn}}{\omega_{mn}^{N+2}}
=
-\frac{1}{N+1}
\sum_{m\ne n}\frac{f_{mn}}{\omega_{mn}^{N+1}}
\left[
(-1)^N\mathcal C^{mn}_{cab}
+\mathcal C^{mn}_{bac}
\right],
\\[4pt]
&\sum_{m\ne n}
\frac{f_{mn}\Delta^b_{mn}
r^a_{nm}r^c_{mn}}{\omega_{mn}^{N+2}}
=
\frac{1}{N+1}
\sum_{m\ne n}\frac{f_{mn}}{\omega_{mn}^{N+1}}
\left[
(-1)^N\mathcal C^{mn}_{cba}
+\mathcal C^{mn}_{abc}
\right].
\end{aligned}
\label{eq:subgap_velocity_reduction}
\end{equation}
The remaining generalized-derivative products
are converted using
$r^{b;a}_{nm}r^c_{mn}
=\mathcal C^{nm}_{cab}$ and
$r^{a;b}_{nm}r^c_{mn}
=\mathcal C^{nm}_{cba}$.
Noting that $\alpha_N(-1)^N=\alpha_N$, we obtain
the general connection coefficient

\begin{equation}
\begin{aligned}
\mathcal G_{N,\mathcal C}^{abc}
=
\frac{q^3}{(N+1)\hbar^2}
\sum_{m\ne n}
\frac{f_{mn}}{\omega_{mn}^{N+1}}
\Big[
&\big((N+1)(-1)^N-\alpha_N\big)
\mathcal C^{mn}_{cab}
+(-1)^N\big((N+1)\beta_N-\gamma_N\big)
\mathcal C^{mn}_{cba}
-\gamma_N\mathcal C^{mn}_{abc}
+(b\leftrightarrow c)
\Big].
\end{aligned}
\label{eq:subgap_connection_general_coefficient}
\end{equation}
This equality holds after the band summation
and Brillouin-zone integration.
The coefficients multiplying the three
Hermitian-connection tensors through fourth order are
\begin{equation}
\begin{array}{c|rrr}
N
& \mathcal C^{mn}_{cab}
& \mathcal C^{mn}_{cba}
& \mathcal C^{mn}_{abc}
\\ \hline
0 & 0  & 0   & 0   \\
1 & -2 & -2  & -2  \\
2 & 2  & 8   & -10 \\
3 & -4 & -22 & -34 \\
4 & 4  & 52  & -98
\end{array}.
\label{eq:subgap_connection_weight_table}
\end{equation}

The band-exchange identity also determines the geometric
sector contributing at each frequency order.
Since $\mathcal C^{nm}_{abc}
=(\mathcal C^{mn}_{abc})^*$, we obtain
\begin{equation}
\sum_{m\ne n}
\frac{f_{mn}}{\omega_{mn}^{N+1}}
\mathcal C^{mn}_{abc}
=
\begin{cases}
\displaystyle
\sum_{m\ne n}
\frac{f_{mn}}{\omega_{mn}^{N+1}}
\operatorname{Re}\mathcal C^{mn}_{abc},
& N\ \text{even},
\\[6pt]
\displaystyle
i\sum_{m\ne n}
\frac{f_{mn}}{\omega_{mn}^{N+1}}
\operatorname{Im}\mathcal C^{mn}_{abc},
& N\ \text{odd}.
\end{cases}
\label{eq:subgap_connection_real_imag}
\end{equation}
Thus, even- and odd-order connection contributions
involve the real (Riemannian) and imaginary
(symplectic) sectors, respectively.
This distinction follows from Hermiticity and band
exchange alone, without imposing $\mathcal{PT}$ symmetry.

The remaining contribution arises from the explicit
three-band processes in the primed commutator of
Eq.~\eqref{eq:SI_final_sigma_total}.
Expanding the frequency-dependent denominators gives,
for $N\geq1$,
\begin{equation}
\begin{split}
\mathcal G_{N,3\mathrm b}^{abc}
=
\frac{iq^3}{\hbar^2}
\sum_{\substack{m\ne n\\\ell\ne m,n}}
\sum_{s=1}^{N}
\frac{2^s r^a_{nm}}{\omega_{mn}^{s}}
\left[
\frac{f_{\ell n}r^b_{m\ell}r^c_{\ell n}}
{\omega_{\ell n}^{N-s+1}}
-
\frac{f_{m\ell}r^c_{m\ell}r^b_{\ell n}}
{\omega_{m\ell}^{N-s+1}}
\right]
+(b\leftrightarrow c).
\end{split}
\label{eq:subgap_three_band_coeff}
\end{equation}
Here $(b\leftrightarrow c)$ adds the exchanged
contribution once. Together with
Eq.~\eqref{eq:subgap_connection_general_coefficient},
this expression determines the complete expansion
coefficient $\mathcal G_N^{abc}$.
In the following, we examine how the combined
coefficients are further constrained by
$\mathcal{PT}$ and crystalline symmetries.

\textit{Static and linear orders.}---
At $N=0$, substituting $\alpha_0=1$ and
$\beta_0=\gamma_0=0$ into
Eq.~\eqref{eq:subgap_connection_general_coefficient}
gives $\mathcal G_{0,\mathcal C}^{abc}=0$.
Since the explicit three-band contribution also
vanishes at this order,
$\mathcal G_0^{abc}=0$, consistent with the
static total-derivative cancellation in
Eq.~\eqref{eq:SHG_static_limit}.

At linear order ($N=1$), the connection
contribution becomes
\begin{equation}
\mathcal G_{1,\mathcal C}^{abc}
=
-\frac{q^3}{\hbar^2}
\sum_{m\ne n}
\frac{f_{mn}}{\omega_{mn}^2}
\left[
\mathcal C^{mn}_{cab}
+\mathcal C^{mn}_{cba}
+\mathcal C^{mn}_{abc}
+(b\leftrightarrow c)
\right].
\label{eq:subgap_linear_connection_unsym}
\end{equation}
The bracket contains all six Cartesian
permutations and equals
$6\mathcal C^{mn}_{(abc)}$.
Using the odd-order band-exchange relation
in Eq.~\eqref{eq:subgap_connection_real_imag},
we obtain
\begin{equation}
\mathcal G_1^{abc}
=
-\frac{6iq^3}{\hbar^2}
\sum_{m\ne n}
\frac{f_{mn}}{\omega_{mn}^2}
\operatorname{Im}\mathcal C^{mn}_{(abc)}
+\mathcal G_{1,3\mathrm b}^{abc}.
\label{eq:subgap_linear_connection}
\end{equation}
At the same order, combining band-exchanged
contributions allows the three-band term
to be expressed in terms of real parts
of closed triple products,
\begin{equation}
\begin{split}
\mathcal G_{1,3\mathrm b}^{abc}
=
\frac{2iq^3}{\hbar^2}
\sum_{\substack{m\ne n\\\ell\ne m,n}}
\frac{1}{\omega_{mn}}
\left[
\frac{f_{\ell n}}{\omega_{\ell n}}
\operatorname{Re}
\left(r^a_{nm}r^b_{m\ell}r^c_{\ell n}\right)
-
\frac{f_{m\ell}}{\omega_{m\ell}}
\operatorname{Re}
\left(r^a_{nm}r^c_{m\ell}r^b_{\ell n}\right)
\right]
+(b\leftrightarrow c).
\end{split}
\label{eq:subgap_linear_three_band}
\end{equation}

For a $\mathcal{PT}$-symmetric system,
the antiunitary $\mathcal{PT}$ operation
leaves crystal momentum unchanged.
The gauge-invariant interband geometric
quantities transform as
\begin{equation}
\begin{aligned}
Q&\xrightarrow{\mathcal{PT}}Q^*,
\\
\mathcal C&\xrightarrow{\mathcal{PT}}\mathcal C^*,
\\
\operatorname{Tr}
\left(r^a_{nm}r^b_{m\ell}r^c_{\ell n}\right)
&\xrightarrow{\mathcal{PT}}
-
\left[
\operatorname{Tr}
\left(r^a_{nm}r^b_{m\ell}r^c_{\ell n}\right)
\right]^*.
\end{aligned}
\label{eq:subgap_PT_geometry}
\end{equation}
For spinful $\mathcal{PT}$-protected multiplets,
these relations apply to the complete
subspace traces, rather than individual
matrix elements in an arbitrary Kramers
basis~\cite{Ahn2022}.

Consequently, the imaginary part of the
Hermitian connection and the real parts
of the closed three-band products vanish
after the $\mathcal{PT}$ symmetry sum.
Both contributions to the linear coefficient
therefore vanish, giving
$\mathcal G_1^{abc}=0$.

This result extends to all odd frequency orders.
In the clean subgap regime, the frequency
denominators are real, and the
$\mathcal{PT}$ transformation properties
in Eq.~\eqref{eq:subgap_PT_geometry}
ensure that the complete SHG conductivity,
including the three-band contribution,
is real. Combined with the physical
reality condition in Eq.~\eqref{eq:subgap_reality},
this requires
\begin{equation}
\mathcal G_{2j+1}^{abc}=0
\qquad (j\geq0),
\end{equation}
so the subgap conductivity takes the form
\begin{equation}
\sigma^{abc}(2\omega;\omega,\omega)
=
\omega^2\mathcal G_2^{abc}
+\omega^4\mathcal G_4^{abc}
+O(\omega^6).
\label{eq:subgap_PT_even}
\end{equation}
Importantly, this selection rule applies
to the complete multiband response and
cannot be inferred from the Hermitian
connection alone.

\textit{Quadratic term.}---
Setting $N=2$ in
Eq.~\eqref{eq:subgap_connection_general_coefficient}
and including the explicit three-band contribution,
we obtain
\begin{equation}
\begin{aligned}
\mathcal G_2^{abc}
=
\frac{q^3}{3\hbar^2}
\sum_{m\ne n}
\frac{f_{mn}}{\omega_{mn}^3}
\operatorname{Re}
\Big[
&2\mathcal C^{mn}_{cab}
+8\mathcal C^{mn}_{cba}
-10\mathcal C^{mn}_{abc}
+(b\leftrightarrow c)
\Big]
+\mathcal G_{2,3\mathrm b}^{abc}.
\end{aligned}
\label{eq:omega^2_final}
\end{equation}
In the two-band limit, the torsion vanishes,
$\mathcal C^{mn}_{abc}=\mathcal C^{mn}_{acb}$,
and the explicit three-band contribution is absent.
The connection terms then combine into
\begin{equation}
\mathcal G_{2,\mathcal{C}}^{abc}
=
\frac{10q^3}{\hbar^2}
\sum_{m\ne n}
\frac{f_{mn}}{\omega_{mn}^3}
\left[
\operatorname{Re}\mathcal C^{mn}_{(abc)}
-\operatorname{Re}\mathcal C^{mn}_{abc}
\right],
\label{eq:subgap_G2_two_band}
\end{equation}
reproducing Eq.~(9)
of the main text. Equation~\eqref{eq:subgap_G2_two_band}
makes the vanishing of the fully symmetric quadratic
response manifest in the two-band limit.

The same cancellation holds for the Hermitian-connection
contribution in the general multiband case, as follows
directly from the full Cartesian symmetrization of
Eq.~\eqref{eq:omega^2_final}:
$\mathcal G_{2,\mathcal C}^{(abc)}=0$.
To establish the corresponding result for the complete
multiband response, we now examine the explicit
three-band contribution.
Setting $N=2$ in
Eq.~\eqref{eq:subgap_three_band_coeff},
we obtain the frequency kernel
\begin{equation}
\begin{aligned}
K_{mn\ell}^{[2]}
={}&
\frac{2}{\omega_{mn}}
\left(
\frac{f_{\ell n}}{\omega_{\ell n}^2}
-\frac{f_{m\ell}}{\omega_{m\ell}^2}
\right)
+
\frac{4}{\omega_{mn}^2}
\left(
\frac{f_{\ell n}}{\omega_{\ell n}}
-\frac{f_{m\ell}}{\omega_{m\ell}}
\right).
\end{aligned}
\label{eq:subgap_quadratic_kernel}
\end{equation}
Using the relations among the band-energy and
occupation differences, this kernel satisfies
the cyclic identity
\begin{equation}
K_{mn\ell}^{[2]}
+K_{n\ell m}^{[2]}
+K_{\ell mn}^{[2]}
=0.
\label{eq:subgap_quadratic_cyclic}
\end{equation}
After full Cartesian symmetrization, the
closed three-band products are invariant under
cyclic permutations of $(m,n,\ell)$,
by cyclicity of the trace.
The corresponding frequency kernels therefore
cancel upon summation over band triples,
yielding
\begin{equation}
\mathcal G_{2,3\mathrm b}^{(abc)}=0,
\qquad
\mathcal G_2^{(abc)}=0.
\label{eq:subgap_G2_fully_symmetric}
\end{equation}
Thus, the fully symmetric quadratic response
vanishes in the complete multiband theory,
independently of $\mathcal{PT}$ symmetry.

For bilayer MnBi$_2$Te$_4$ with out-of-plane
N\'eel order, the $C_{3z}$ and
$\mathcal M_x\mathcal T$ symmetries constrain
the in-plane SHG tensor through
Eqs.~(13)
and~(14).
In particular,
$\mathcal G_2^{xyy}
=\mathcal G_2^{yxy}
=\mathcal G_2^{yyx}
=-\mathcal G_2^{xxx}$.
Since these allowed components are fully
symmetric under Cartesian index permutations,
$\mathcal G_2^{(abc)}=0$ forces all in-plane
quadratic coefficients to vanish.
In contrast, a general $\mathcal{PT}$-symmetric
insulator without these additional crystalline
constraints may retain a nonzero quadratic
response with a vanishing fully symmetric part [shown in Fig.~\ref{fig:small_angle_activation}(a)].

\textit{Quartic term.}---
Setting $N=4$ in
Eq.~\eqref{eq:subgap_connection_general_coefficient}
and including the explicit three-band contribution,
we obtain
\begin{equation}
\begin{aligned}
\mathcal G_4^{abc}
=
\frac{q^3}{5\hbar^2}
\sum_{m\ne n}
\frac{f_{mn}}{\omega_{mn}^5}
\operatorname{Re}
\Big[
&4\mathcal C^{mn}_{cab}
+52\mathcal C^{mn}_{cba}
-98\mathcal C^{mn}_{abc}
+(b\leftrightarrow c)
\Big]
+\mathcal G_{4,3\mathrm b}^{abc}.
\end{aligned}
\label{eq:omega^4_final}
\end{equation}
The three-band contribution is given by
Eq.~\eqref{eq:subgap_three_band_coeff}
with $N=4$.

In the two-band limit, the torsion vanishes,
$\mathcal C^{mn}_{abc}=\mathcal C^{mn}_{acb}$,
and the explicit three-band contribution is absent.
The quartic coefficient then reduces to
\begin{equation}
\mathcal G_{4,\mathcal C}^{abc}
=
\frac{84q^3}{5\hbar^2}
\sum_{m\ne n}
\frac{f_{mn}}{\omega_{mn}^5}
\left[
2\operatorname{Re}\mathcal C^{mn}_{(abc)}
-
3\operatorname{Re}\mathcal C^{mn}_{abc}
\right].
\label{eq:subgap_G4_two_band}
\end{equation}
In particular, the longitudinal connection contribution is
\begin{equation}
\mathcal G_{4,\mathcal C}^{xxx}
=
-\frac{84q^3}{5\hbar^2}
\sum_{m\ne n}
\frac{f_{mn}}{\omega_{mn}^5}
\operatorname{Re}\mathcal C^{mn}_{xxx},
\end{equation}
which need not vanish. This expression for the
connection contribution also holds in the
general multiband case.

For bilayer MnBi$_2$Te$_4$ with out-of-plane
N\'eel order, the vanishing static and quadratic
coefficients, together with the absence of odd
frequency orders enforced by $\mathcal{PT}$ symmetry,
imply that the leading in-plane SHG response is
quartic in frequency:
\begin{equation}
\sigma^{abc}(2\omega;\omega,\omega)
=
\omega^4\mathcal G_4^{abc}
+O(\omega^6),
\qquad a,b,c\in\{x,y\}.
\label{eq:subgap_bilayer_omega4}
\end{equation}
The quartic coefficient contains energy-weighted
Hermitian-connection contributions as well as genuine
three-band processes, both retained in the eight-band
model through Eq.~\eqref{eq:omega^4_final}.
While $\mathcal{PT}$ symmetry eliminates odd powers
of frequency, the absence of the quadratic
in-plane response additionally relies on the
crystalline symmetries of the out-of-plane
magnetic configuration.
We next examine how rotating the N\'eel vector
relaxes these constraints.

\subsubsection{Small-angle expansion for N\'eel-vector rotation}

\begin{figure}[htbp]
    \centering
    \includegraphics[width=0.6\linewidth]{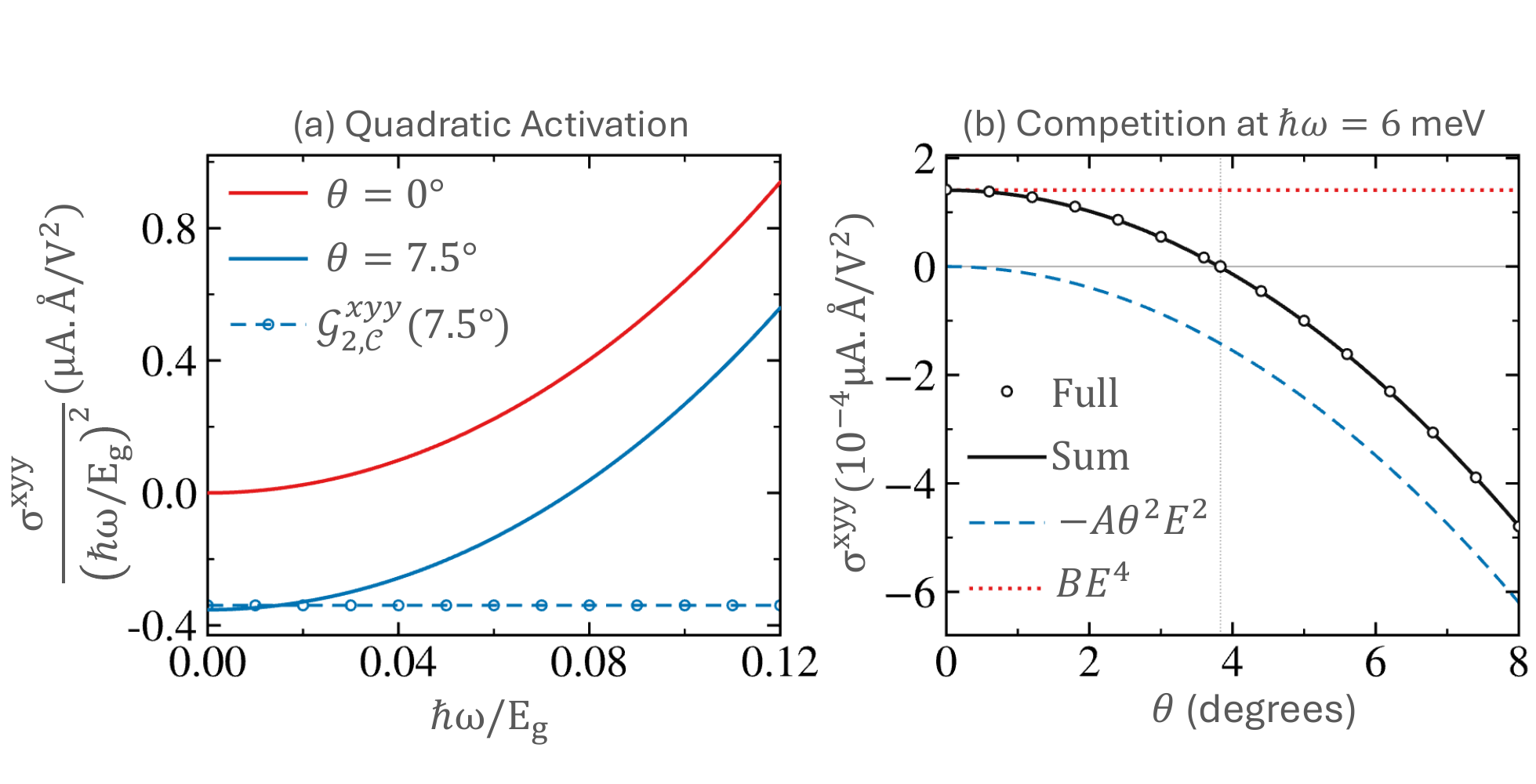}
    \caption{\textbf{Quadratic activation and competition of subgap SHG
    upon N\'eel-vector rotation in bilayer MnBi$_2$Te$_4$
    at $m_0=-110$ meV.}
    (a) Normalized conductivity
    $\sigma^{xyy}/(\hbar\omega/E_g)^2$
    for $\theta=0^\circ$ (red) and
    $\theta=7.5^\circ$ (blue), where
    $E_g$ is the electronic gap at each angle.
    The dashed blue line with open circles shows
    the normalized Hermitian-connection contribution,
    $(E_g/\hbar)^2\mathcal G_{2,\mathcal C}^{xyy}$,
    at $\theta=7.5^\circ$.
    (b) Angular dependence of $\sigma^{xyy}$
    at $\hbar\omega=6$ meV.
    The full numerical response (open circles)
    is compared with the small-angle approximation
    (solid black), consisting of the quadratic
    $-A\theta^2E^2$ (dashed blue) and quartic
    $BE^4$ (dotted red) contributions,
    with $E=\hbar\omega$.
    }
    \label{fig:small_angle_activation}
\end{figure}

Consider a rotation of the N\'eel vector in the $xz$ plane,
$\mathbf n(\theta)=(\sin\theta,0,\cos\theta)$.
The $C_{2x}$ operation exchanges the two layers
and transforms the staggered N\'eel vector as
$(n_x,n_y,n_z)\mapsto(-n_x,n_y,n_z)$,
mapping $\mathbf n(\theta)$ to $\mathbf n(-\theta)$.
Since $j_x$ and $E_y^2$ are invariant under this
operation, the conductivity satisfies
\begin{equation}
\sigma^{xyy}(-\theta,\omega)
=
\sigma^{xyy}(\theta,\omega).
\end{equation}
The frequency-expansion coefficients are therefore
even functions of $\theta$.
Together with $\mathcal G_2^{xyy}(0)=0$,
this gives
\begin{align}
\mathcal G_2^{xyy}(\theta)
&=
\frac{1}{2}\mathcal G_2^{xyy\,\prime\prime}(0)
\theta^2+O(\theta^4),
\\
\mathcal G_4^{xyy}(\theta)
&=
\mathcal G_4^{xyy}(0)+O(\theta^2).
\end{align}
Expressing the response in terms of the incident
photon energy $E=\hbar\omega$, we obtain
\begin{align}
\sigma^{xyy}(\theta,E)
&=
-A\theta^2E^2+BE^4
+O(\theta^4E^2,\theta^2E^4,E^6),
\\
A&=
-\frac{\mathcal G_2^{xyy\,\prime\prime}(0)}
{2\hbar^2},
\qquad
B=\frac{\mathcal G_4^{xyy}(0)}{\hbar^4}.
\end{align}
Here $\theta$ is measured in radians.
Both coefficients include the Hermitian-connection
and genuine three-band contributions.

When $AB>0$, the quadratic and quartic terms have
opposite signs, producing a zero crossing at
\begin{equation}
\theta_*(E)\simeq E\sqrt{\frac{B}{A}}.
\label{eq:small_angle_crossing}
\end{equation}
Thus, the crossing angle scales linearly with
photon energy in the low-frequency limit.
The existence of such a crossing depends on the
relative signs of the coefficients and is not
enforced by symmetry.

Figure~\ref{fig:small_angle_activation} illustrates
these results using the eight-band model.
In Fig.~\ref{fig:small_angle_activation}(a),
the normalized conductivity
$\sigma^{xyy}/(\hbar\omega/E_g)^2$
vanishes quadratically with frequency at
$\theta=0^\circ$, consistent with the quartic
onset shown in Fig.~3 of the main text.
At finite tilt, the activated quadratic response
produces a nonzero low-frequency limit,
\begin{equation}
\lim_{\omega\to0}
\frac{\sigma^{xyy}(\theta,\omega)}
{(\hbar\omega/E_g(\theta))^2}
=
\frac{E_g^2(\theta)}{\hbar^2}
\mathcal G_2^{xyy}(\theta).
\end{equation}
The Hermitian-connection contribution closely
approximates this limit, with a smaller additional
contribution from the genuine three-band processes.

Figure~\ref{fig:small_angle_activation}(b)
shows the angular dependence at
$\hbar\omega=6$ meV.
The negative quadratic contribution competes
with the positive quartic contribution,
causing the SHG conductivity to change sign
as the N\'eel vector rotates.
Their sum closely reproduces the full numerical
response, with a predicted zero crossing at
$\theta_*\simeq3.81^\circ$, compared with
$3.83^\circ$ from the full calculation.
This agreement confirms that the sign reversal
arises from the competition between the two
frequency orders.
For other rotation planes, the even-in-$\theta$
constraint need not apply, and linear angular
terms may become symmetry allowed.

\section{Hamiltonian parameters and warping term}
\label{sec:Hamiltonian_parameters}

We describe bilayer (2-SL) MnBi$_2$Te$_4$ using
an eight-band Hamiltonian~\cite{Xu2019,Cheng2024},
\begin{equation}
H_0(\boldsymbol{k})=
\begin{pmatrix}
h_1(\boldsymbol{k}) & -T_{12}\\
-T_{12}^{\dagger} & h_2(\boldsymbol{k})
\end{pmatrix},
\label{eq:MBT_Hamiltonian}
\end{equation}
where $\boldsymbol{k}=(k_x,k_y)$ is the in-plane
momentum measured from $\Gamma$.
The diagonal blocks describe the individual
septuple layers,
\begin{equation}
h_i(\boldsymbol{k})
=
h_{\rm o}(\boldsymbol{k})
+h_{\rm w}(\boldsymbol{k})
+\eta_i h_{\rm m}(\boldsymbol{n}),
\qquad
\eta_1=+1,\quad\eta_2=-1.
\label{eq:MBT_layer_blocks}
\end{equation}
The opposite exchange signs describe the
antiferromagnetic alignment, while $T_{12}$
accounts for interlayer hopping.

Within each layer, we use the basis
\begin{equation}
\left(
\ket{p_{z,\mathrm{Bi}}^{+},\uparrow},
\ket{p_{z,\mathrm{Te}}^{-},\uparrow},
\ket{p_{z,\mathrm{Bi}}^{+},\downarrow},
\ket{p_{z,\mathrm{Te}}^{-},\downarrow}
\right)^T.
\label{eq:MBT_basis}
\end{equation}
The orbital Hamiltonian is
\begin{equation}
h_{\rm o}(\boldsymbol{k})=
\begin{pmatrix}
d_0(\boldsymbol{k}) & 0 & 0 & A_2 k_-\\
0 & -d_0(\boldsymbol{k}) & A_2 k_- & 0\\
0 & A_2 k_+ & d_0(\boldsymbol{k}) & 0\\
A_2 k_+ & 0 & 0 & -d_0(\boldsymbol{k})
\end{pmatrix},
\label{eq:MBT_orbital}
\end{equation}
with
\begin{equation}
d_0(\boldsymbol{k})
=
m_0+2B_1+B_2(k_x^2+k_y^2),
\qquad
k_{\pm}=k_x\pm ik_y.
\label{eq:MBT_mass}
\end{equation}
The common scalar dispersion
$\epsilon_0(\boldsymbol{k})\mathbb{1}_4$
is omitted from the numerical Hamiltonian.
Consequently, the parameters $C$ and $D_2$
associated with this term do not enter
the calculations, whereas $D_1$ is retained
in the interlayer hopping below.

The cubic warping term is
\begin{equation}
\begin{aligned}
h_{\rm w}(\boldsymbol{k})
&=
-i\lambda_w(k_+^3-k_-^3)
\,\sigma_z\otimes\tau_x\\
&=
2\lambda_w k_y(3k_x^2-k_y^2)
\,\sigma_z\otimes\tau_x,
\end{aligned}
\label{eq:MBT_warping}
\end{equation}
where $\sigma_\alpha$ and $\tau_\alpha$ are
Pauli matrices acting on the spin and orbital
degrees of freedom, respectively.
This term reduces continuous in-plane
rotational symmetry to $C_{3z}$.
With momentum expressed in $\mathrm{nm}^{-1}$,
the warping parameter $\lambda_w$ has units
of $\mathrm{eV\,nm^3}$.

The unit N\'eel vector is parametrized as
\begin{equation}
\boldsymbol{n}
=
(\sin\theta\cos\phi,\,
 \sin\theta\sin\phi,\,
 \cos\theta),
\qquad
n_{\pm}=n_x\pm in_y.
\label{eq:MBT_Neel_vector}
\end{equation}
The exchange Hamiltonian in the first layer is
\begin{equation}
h_{\rm m}(\boldsymbol{n})
=
\gamma_{\rm ex}
\begin{pmatrix}
M_1n_z & 0 & M_1n_- & 0\\
0 & M_2n_z & 0 & M_2n_-\\
M_1n_+ & 0 & -M_1n_z & 0\\
0 & M_2n_+ & 0 & -M_2n_z
\end{pmatrix},
\label{eq:MBT_exchange}
\end{equation}
with the opposite sign in the second layer.
Here $\gamma_{\rm ex}$ sets the exchange
energy scale, while $M_1$ and $M_2$ are
dimensionless orbital-dependent exchange
weights. For out-of-plane N\'eel order,
\begin{equation}
h_{\rm m}(\hat{\boldsymbol{z}})
=
\gamma_{\rm ex}
\operatorname{diag}(M_1,M_2,-M_1,-M_2).
\end{equation}
The angular calculations use $\phi=0$,
corresponding to rotation of the N\'eel
vector in the $xz$ plane.

The interlayer hopping matrix is
\begin{equation}
T_{12}=
\begin{pmatrix}
D_1+B_1 & iA_1 & 0 & 0\\
iA_1 & D_1-B_1 & 0 & 0\\
0 & 0 & D_1+B_1 & -iA_1\\
0 & 0 & -iA_1 & D_1-B_1
\end{pmatrix}.
\label{eq:MBT_hopping}
\end{equation}
Here $A_1$, $B_1$, and $D_1$ have units
of energy. Following the convention of
Tang and Cheng~\cite{Cheng2024}, we report
the dimensional combinations $A_1a_z$,
$B_1a_z^2$, and $D_1a_z^2$ in
Table~\ref{tab:MBT_parameters}, where
$a_z$ is the spacing between adjacent
septuple layers.

The unwarped Hamiltonian and its
band-structure parameters are adapted from
Tang and Cheng~\cite{Cheng2024}, based on
earlier first-principles
descriptions~\cite{Xu2019}.
We use $a_z=1.36\,\mathrm{nm}$, consistent
with structural measurements~\cite{Gong2019}.
The reference band-inversion parameter is
$m_0=-110\,\mathrm{meV}$, which is varied
as indicated in the corresponding figures.

For the cubic warping strength, we adopt
$\lambda_w=0.040\,\mathrm{eV\,nm^3}$,
motivated by the spin-resolved photoemission
study of Han \textit{et al.}~\cite{Han2025}.
Their effective surface-state Hamiltonian
contains a warping term
$\gamma(k_+^3+k_-^3)\sigma_z$, with a fitted
magnitude $|\gamma|=39\,\mathrm{eV\,\AA^3}
=0.039\,\mathrm{eV\,nm^3}$ for MnBi$_2$Te$_4$.
The two cubic functions have equivalent
angular dependence up to an in-plane
rotation of the coordinate axes.
However, the experimental fit describes
a surface-state doublet rather than the
orbital-resolved bilayer Hamiltonian used
here. We therefore use the experimental
result to motivate a representative warping
scale, rather than as a direct determination
of the bilayer warping parameter.
The model parameters are summarized in
Table~\ref{tab:MBT_parameters}.

\begin{table}[tb]
\centering
\caption{
Parameters used in the bilayer Hamiltonian.
The dimensional combinations $A_1a_z$,
$B_1a_z^2$, and $D_1a_z^2$ follow the
convention of Ref.~\cite{Cheng2024}.
Values are rounded to three significant
digits for presentation, while the
calculations use unrounded parameters.
The reference mass is $m_0=-110\,\mathrm{meV}$;
other values are specified in the
corresponding figures.
}
\label{tab:MBT_parameters}
\begin{tabular}{ccc}
\hline\hline
Parameter & Value & Unit\\
\hline
$a_z$             & $1.36$   & $\mathrm{nm}$\\
$A_1a_z$          & $0.135$  & $\mathrm{eV\,nm}$\\
$A_2$             & $0.320$  & $\mathrm{eV\,nm}$\\
$B_1a_z^2$        & $0.119$  & $\mathrm{eV\,nm^2}$\\
$B_2$             & $0.0940$ & $\mathrm{eV\,nm^2}$\\
$D_1a_z^2$        & $0.0272$ & $\mathrm{eV\,nm^2}$\\
$\gamma_{\rm ex}$ & $0.0550$ & $\mathrm{eV}$\\
$M_1$             & $1.0$    & dimensionless\\
$M_2$             & $1.5$    & dimensionless\\
$\lambda_w$       & $0.0400$ & $\mathrm{eV\,nm^3}$\\
$m_0$ (reference) & $-0.110$ & $\mathrm{eV}$\\
\hline\hline
\end{tabular}
\end{table}

Unless otherwise specified, the continuum
momentum integrals are evaluated over the
circular domain $|\boldsymbol{k}|\leq k_c$,
with $k_c=2\,\mathrm{nm}^{-1}
=0.2\,\text{\AA}^{-1}$.
The four lowest bands are occupied in the
insulating calculations.

\begin{figure}[htbp]
    \centering
    \includegraphics[width=0.75\linewidth]
    {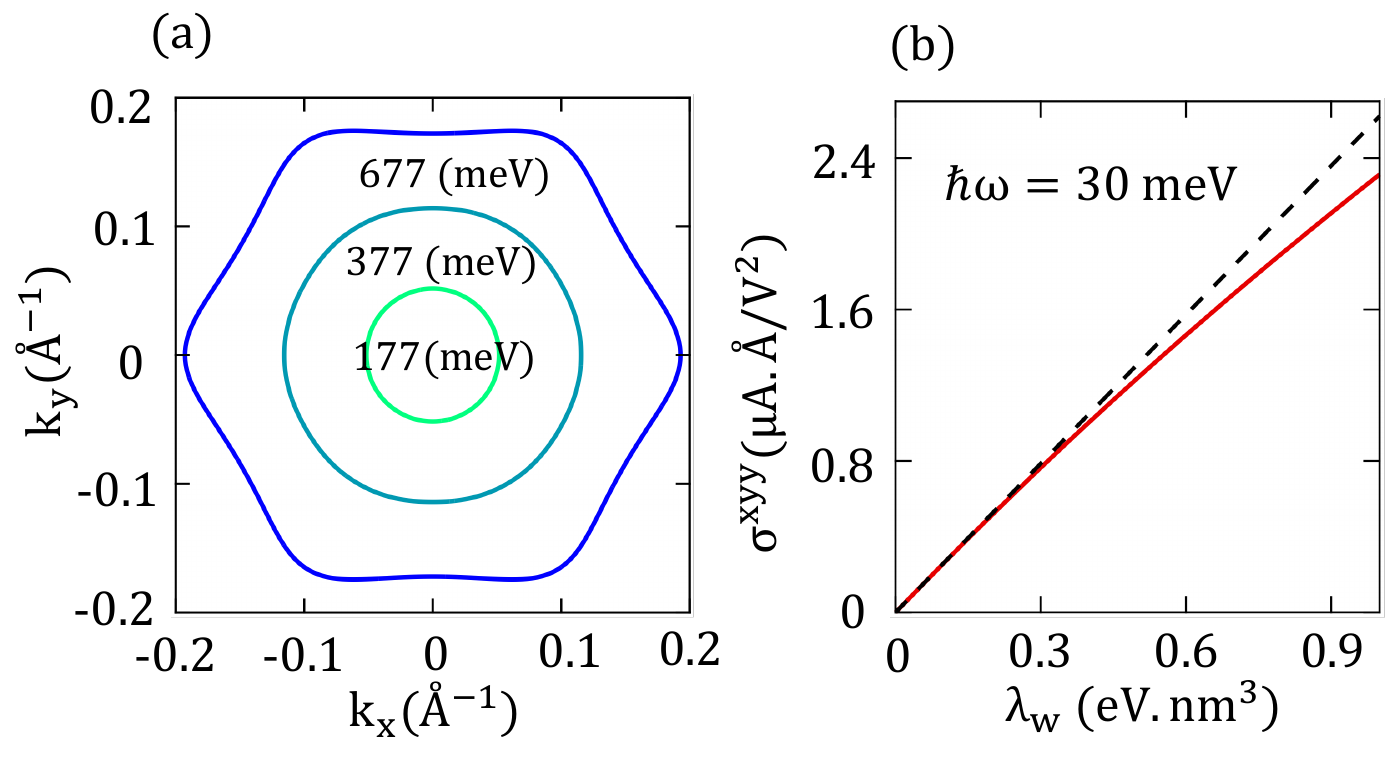}
    \caption{
    Effect of cubic warping on the electronic
    structure and subgap SHG response of
    bilayer MnBi$_2$Te$_4$.
    (a) Constant-energy contours at
    177, 377, and 677 meV, illustrating
    the increasing threefold distortion
    at higher energies.
    (b) Dependence of the in-plane SHG
    response on the warping strength
    $\lambda_w$ at $\hbar\omega=30$ meV.
    The solid red curve shows the numerical
    result, while the black dashed line
    indicates a linear reference trend.
    The response vanishes at $\lambda_w=0$
    and remains approximately linear over
    a broad parameter range.
    Both panels use the parameters in
    Table~\ref{tab:MBT_parameters}, with
    $m_0=-110$ meV and out-of-plane
    N\'eel order. The warping strength is
    varied only in (b).
    }
    \label{fig:figure05}
\end{figure}

To examine the effect of cubic warping,
Fig.~\ref{fig:figure05}(a) shows
representative constant-energy contours.
Since $h_{\rm w}$ scales as
$\lambda_w k^3$, its influence increases
with momentum. The lower-energy contour
is nearly circular, whereas higher-energy
contours exhibit increasingly pronounced
threefold distortion.

The warping term also determines whether
the in-plane SHG response is symmetry
allowed. For out-of-plane N\'eel order,
setting $\lambda_w=0$ restores continuous
in-plane rotational symmetry. In particular,
a rotation by $\pi$ reverses $j_x$ while
leaving $E_y^2$ invariant, requiring
$\sigma^{xyy}=0$.
Finite warping reduces the symmetry to
$C_{3z}$, allowing the nonzero in-plane
tensor components discussed in the
main text.

Figure~\ref{fig:figure05}(b) shows the
dependence of the SHG response on
$\lambda_w$ at $\hbar\omega=30$ meV.
The response magnitude grows approximately
linearly with the warping strength over
a broad parameter range, with noticeable
deviations occurring only at relatively
large values of $\lambda_w$.
Our adopted value,
$\lambda_w=0.040\,\mathrm{eV\,nm^3}$,
lies well within this nearly linear regime.
These results show that the warping term
provides the necessary symmetry breaking
for the in-plane SHG response, while its
magnitude varies smoothly with the
phenomenological warping strength.

\section{Layer Chern number}
\label{sec:layer_chern}

To characterize the layer-resolved topology of the occupied
bands, we adapt the projected-spin construction of
Prodan~\cite{Prodan2009} to the layer degree of freedom.
The occupied-band projector is defined as
\begin{equation}
P(\mathbf{k})
=
\sum_{n\in\mathrm{occ}}
|u_{\mathbf{k}n}\rangle
\langle u_{\mathbf{k}n}|.
\label{eq:projector}
\end{equation}
In the layer-spin-orbital basis of Eq.~(11) in the main text,
the layer-pseudospin operator is
\begin{equation}
\mathcal O_\ell
=
\mu_z\otimes\sigma_0\otimes\tau_0
=
\begin{pmatrix}
I_4 & 0\\
0 & -I_4
\end{pmatrix},
\label{eq:layer_operator}
\end{equation}
whose eigenvalues distinguish the two septuple layers.
Since interlayer hybridization generally mixes the layer
degrees of freedom, we consider the projected operator
\begin{equation}
Q_\ell(\mathbf{k})
=
P(\mathbf{k})\mathcal O_\ell P(\mathbf{k}).
\label{eq:projected_layer_operator}
\end{equation}
Its eigenstates and eigenvalues within the occupied
subspace satisfy
\begin{equation}
Q_\ell(\mathbf{k})|\phi_{\mathbf{k}m}\rangle
=
\xi_{\mathbf{k}m}|\phi_{\mathbf{k}m}\rangle,
\qquad
P(\mathbf{k})|\phi_{\mathbf{k}m}\rangle
=
|\phi_{\mathbf{k}m}\rangle.
\label{eq:spectralenergyequation}
\end{equation}
Numerically, we diagonalize
$U_{\mathrm{occ}}^\dagger\mathcal O_\ell U_{\mathrm{occ}}$,
where the columns of $U_{\mathrm{occ}}$ contain the
occupied Bloch eigenstates. This restriction excludes
the trivial zero eigenvalues associated with the
unoccupied subspace.

The dimensionless eigenvalues
$\xi_{\mathbf{k}m}\in[-1,1]$ constitute the
\emph{projected layer-pseudospin spectrum}, which is
invariant under unitary rotations of the occupied Bloch
basis. In the $\mathcal{PT}$-symmetric bilayer,
$\mathcal{PT}$ preserves $P(\mathbf{k})$ while reversing
$\mathcal O_\ell$. Consequently, the projected-layer
eigenvalues occur in pairs $\pm\xi$ at each momentum.
We define the spectral gap separating the positive and
negative sectors as
\begin{equation}
E_{\rm sg}
=
2\min_{\mathbf{k}\in\mathrm{BZ},\,m}
|\xi_{\mathbf{k}m}|.
\label{eq:layer_spectral_gap}
\end{equation}
Unlike the electronic excitation gap $E_{\rm g}$,
$E_{\rm sg}$ is dimensionless.

When $E_{\rm sg}>0$, the occupied manifold separates
into positive and negative projected-layer sectors,
\begin{equation}
P=P_++P_-,
\qquad
P_\pm(\mathbf{k})
=
\sum_{m:\,\pm\xi_{\mathbf{k}m}>0}
|\phi_{\mathbf{k}m}\rangle
\langle\phi_{\mathbf{k}m}|.
\label{eq:layer_sector_projectors}
\end{equation}
These projectors remain well defined despite interlayer
hybridization. Using the Berry-connection convention
$A_a=i\langle u|\partial_a u\rangle$, the Chern number
of each sector is
\begin{equation}
C_\pm
=
\frac{i}{2\pi}
\int_{\mathrm{BZ}} d^2\mathbf{k}\,
\operatorname{Tr}
\left\{
P_\pm
\left[
\partial_{k_x}P_\pm,
\partial_{k_y}P_\pm
\right]
\right\}.
\label{eq:layer_sector_chern}
\end{equation}
These integers remain invariant under continuous
deformations that preserve both the electronic gap
and the projected-layer spectral gap.

The total and layer Chern numbers are defined as
\begin{equation}
C=C_++C_-,
\qquad
C_\ell=\frac{C_+-C_-}{2}.
\label{eq:layer_chern_definition}
\end{equation}
Because $\mathcal{PT}$ exchanges the two sectors
and reverses their Berry curvatures,
$C_-=-C_+$. Thus, the total Chern number vanishes,
$C=0$, while the layer Chern number
$C_\ell=C_+$ can remain nonzero.
Importantly, $C_\ell$ can change when the
projected-layer spectral gap closes, even if the
electronic gap remains open.

\begin{figure}
    \centering
    \includegraphics[width=0.8\linewidth]
    {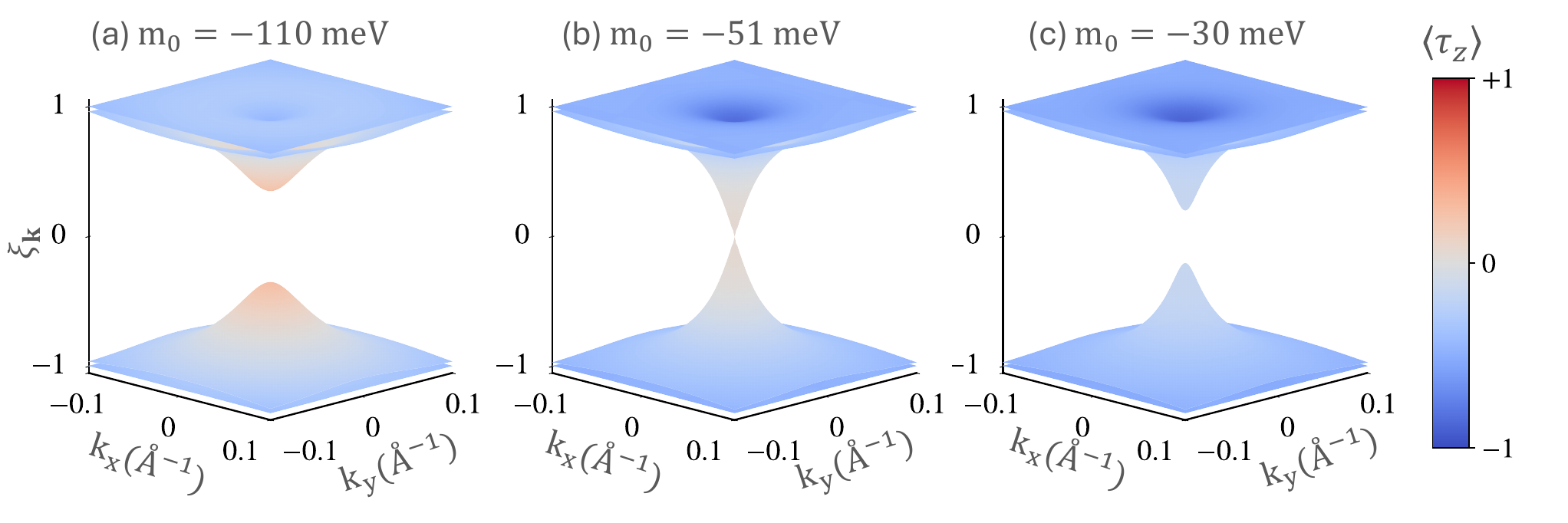}
    \caption{
    \textbf{Projected-layer pseudospin spectrum and
    orbital polarization in bilayer MnBi$_2$Te$_4$.}
    The four surfaces represent the dimensionless
    eigenvalues $\xi_{\mathbf{k}m}$ of
    $Q_\ell(\mathbf{k})=P(\mathbf{k})
    \mathcal O_\ell P(\mathbf{k})$
    within the occupied subspace, for
    (a) $m_0=-110~\mathrm{meV}$ ($C_\ell=-1$),
    (b) $m_0=-51~\mathrm{meV}$ (transition), and
    (c) $m_0=-30~\mathrm{meV}$ ($C_\ell=0$).
    The N\'eel vector is oriented along $z$.
    Colors indicate the orbital polarization
    $\langle\tau_z\rangle_m
    =\langle\phi_{\mathbf{k}m}|
    \tau_z|\phi_{\mathbf{k}m}\rangle$,
    with positive (negative) values corresponding
    to predominantly Bi-like (Te-like) character.
    The nontrivial phase exhibits inverted orbital
    character near $\Gamma$, while the projected-layer
    gap closes at the transition and reopens in the
    trivial phase.
    }
    \label{fig:layer_pseudospectrum}
\end{figure}

Figure~\ref{fig:layer_pseudospectrum} illustrates
the evolution of the projected-layer spectrum and
orbital character across the topological transition.
At $m_0=-110~\mathrm{meV}$, the occupied states exhibit
pronounced orbital mixing and inversion near $\Gamma$,
consistent with the nontrivial layer Chern number
$C_\ell=-1$. As $m_0$ approaches $-51~\mathrm{meV}$,
the projected-layer gap closes at $\Gamma$, allowing
the layer Chern number to change without closing the
electronic gap. At $m_0=-30~\mathrm{meV}$, the
projected-layer gap reopens in the trivial phase
($C_\ell=0$), where the orbital inversion is absent.

As shown in Fig.~(2) of the main text,
the projected-layer gap closes at the boundaries
between the $C_\ell=-1$ and $C_\ell=0$ phases,
while the electronic gap remains finite.
The intermediate phase therefore exhibits nontrivial
layer-resolved topology despite its vanishing total
Chern number. The enhanced subgap SHG observed in
this phase is associated with the underlying interband
geometry, rather than being determined by the
topological invariant alone.

\bibliography{references}{}